\documentclass[11pt,a4paper,nofootinbib]{revtex4}
\usepackage[utf8]{inputenc}
\usepackage{graphicx,color}
\usepackage{fullpage}
\usepackage{eso-pic}
\usepackage{enumitem}
\usepackage{pdfpages}
\usepackage{wrapfig}

\usepackage[english]{babel}

\usepackage{contour}
\usepackage{ulem}

\contourlength{0.8pt}
\newcommand{\myuline}[1]{%
  \uline{\phantom{#1}}%
  \llap{\contour{white}{#1}}%
}

\usepackage{titlesec}
\titlespacing*{\section}
{0pt}{1.8ex plus 1ex minus .2ex}{0.5ex plus .2ex}
\titlespacing*{\subsection}
{0pt}{1.5ex plus 1ex minus .2ex}{0.3ex plus .2ex}

\usepackage{amsmath}
\usepackage{amssymb}

\renewcommand{\thesection}{\arabic{section}}
\renewcommand{\thesubsection}{\thesection.\arabic{subsection}}
\renewcommand{\thesubsubsection}{\thesubsection.\arabic{subsubsection}}
\usepackage[utf8]{inputenc}

\usepackage{xspace}

\newcommand{\PIXIE}{{\it PIXIE}\xspace}
\newcommand{\SPIXIE}{{\it Super-PIXIE}\xspace}
\newcommand{\PRISM}{{\it PRISM}\xspace}
\newcommand{\Planck}{{\it Planck}\xspace}
\newcommand{\CORE}{{\it CORE}\xspace}
\newcommand{\PICO}{{\it PICO}\xspace}
\newcommand{\captiontext}{\small \baselineskip=0.95\baselineskip}

\usepackage{geometry}
\newgeometry{vmargin={19mm}, hmargin={20mm,20mm}} 

\begin{document}

\thispagestyle{empty}

\includepdf{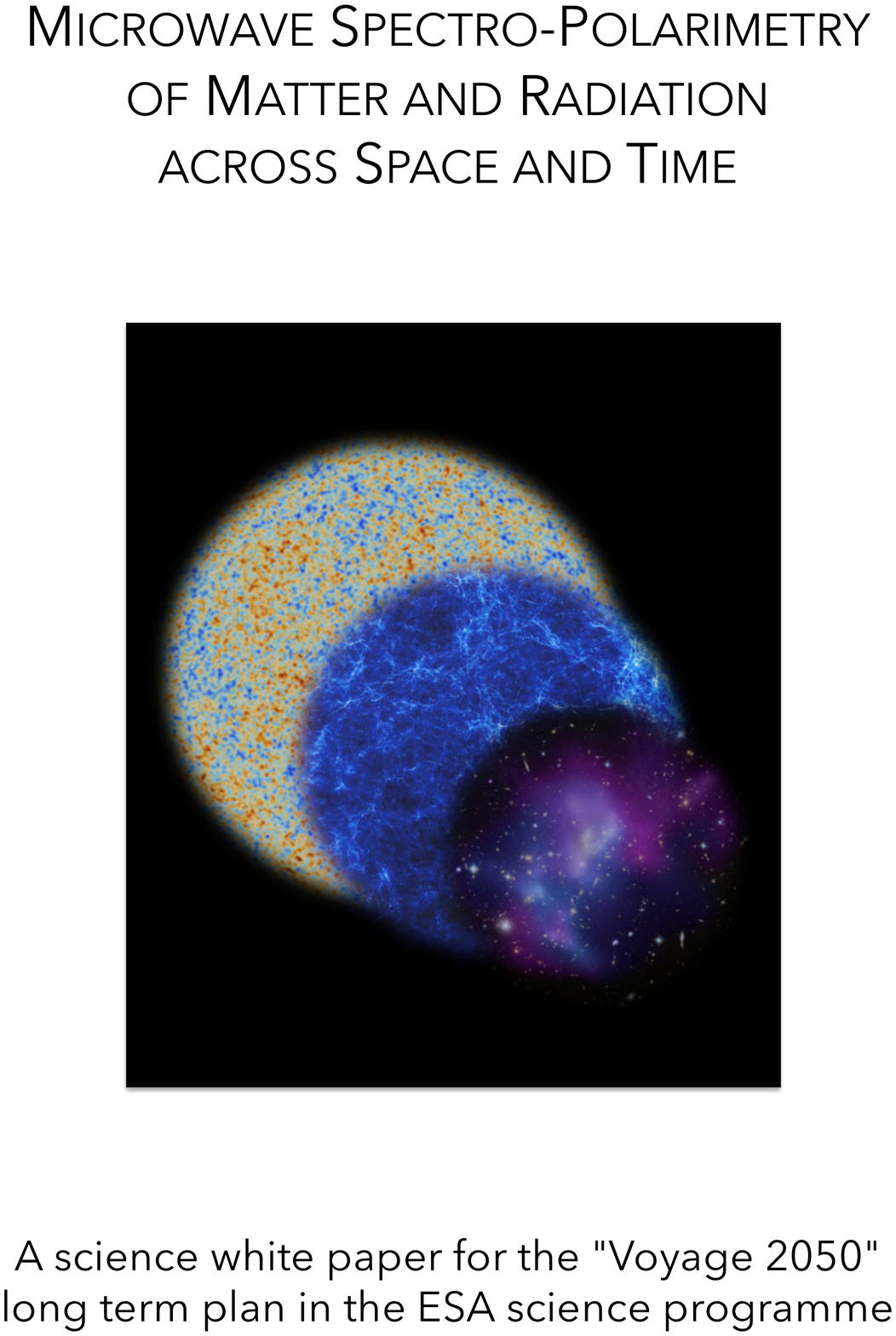}




\vspace{3mm}

\noindent{\small {\bf AUTHORS}\vspace{1mm}
\\
Jacques Delabrouille$^{1,2}$, 
Maximilian~H. Abitbol$^{3}$, 
Nabila Aghanim$^{4}$, 
Yacine Ali-Ha\"imoud$^{5}$, 
David Alonso$^{3,6}$,
Marcelo Alvarez$^{7,8}$, 
Anthony~J. Banday$^{9}$,
James~G. Bartlett$^{1,10}$,
Jochem Baselmans$^{11,12}$,
Kaustuv Basu$^{13}$, 
Nicholas Battaglia$^{14}$,
Jos\'e~Ram\'on Bermejo~Climent$^{15}$,
Jos\'e~L. Bernal$^{16}$, 
Matthieu B\'ethermin$^{17}$,
Boris Bolliet$^{18}$,
Matteo Bonato$^{19,20}$,
Fran\c{c}ois~R. Bouchet$^{21}$,
Patrick~C. Breysse$^{22}$,
Carlo Burigana$^{19}$,
Zhen-Yi~Cai$^{23,24}$,
Jens Chluba$^{18}$,
Eugene Churazov$^{25,26}$,
Helmut Dannerbauer$^{27}$,
Paolo De~Bernardis$^{28,29}$,
Gianfranco De~Zotti$^{20}$,
Eleonora Di~Valentino$^{18}$,
Emanuela Dimastrogiovanni$^{30}$,
Akira Endo$^{11,31}$,
Jens Erler$^{13}$,
Simone Ferraro$^{8,7}$,
Fabio Finelli$^{15}$,
Dale Fixsen$^{32}$,
Shaul Hanany$^{33}$,
Luke Hart$^{18}$,
Carlos Hern\'andez-Monteagudo$^{34}$,
J.~Colin Hill$^{35,36}$,
Selim~C. Hotinli$^{37}$,
Kenichi Karatsu$^{11,12}$,
Kirit Karkare$^{38}$,
Garrett~K. Keating$^{39}$,
Ildar Khabibullin$^{25,26}$,
Alan Kogut$^{40}$,
Kazunori Kohri$^{41}$,
Ely~D. Kovetz$^{42}$,
Guilaine Lagache$^{17}$,
Julien Lesgourgues$^{43}$,
Mathew Madhavacheril$^{44}$,
Bruno Maffei$^{4}$,
Nazzareno Mandolesi$^{45,46}$,
Carlos Martins$^{47,48}$,
Silvia Masi$^{28,29}$,
John Mather$^{40}$,
Jean-Baptiste Melin$^{2}$,
Azadeh Moradinezhad Dizgah$^{49,50}$,
Tony Mroczkowski$^{51}$,
Suvodip Mukherjee$^{21}$,
Daisuke Nagai$^{52}$,
Mattia Negrello$^{6}$,
Nathalie Palanque-Delabrouille$^{2}$,
Daniela Paoletti$^{15}$,
Subodh~P. Patil$^{53}$,
Francesco Piacentini$^{28,29}$,
Srinivasan Raghunathan$^{54}$,
Andrea Ravenni$^{18}$,
Mathieu Remazeilles$^{18}$,
Vincent Rev\'eret$^{2}$,
Louis Rodriguez$^{2}$,
Aditya Rotti$^{18}$,
Jose-Alberto Rubi\~no~Martin$^{27,55}$,
Jack Sayers$^{56}$,
Douglas Scott$^{57}$,
Joseph Silk$^{58,21,59}$,
Marta Silva$^{60}$,
Tarun Souradeep$^{61}$,
Naonori Sugiyama$^{62}$,
Rashid Sunyaev$^{25,26,35}$,
Eric~R. Switzer$^{40}$,
Andrea Tartari$^{63}$,
Tiziana Trombetti$^{19}$,
\'I\~{n}igo Zubeldia$^{64,65}$.
}
\thispagestyle{empty}

\vspace{3mm}
\noindent
{\scriptsize
%
$^{1}$ Laboratoire Astroparticule et Cosmologie (APC), CNRS/IN2P3, Universit\'e Paris Diderot, 75205 Paris Cedex 13, France
\\[-1mm]
%
$^{2}$ IRFU, CEA, Universit\'e Paris Saclay, 91191 Gif-sur-Yvette, France 
\\[-1mm]
%
$^3$ University of Oxford, Denys Wilkinson Building, Keble Road, Oxford, OX1 3RH, UK
\\[-1mm]
%
$^4$ Institut d'Astrophysique Spatiale (IAS), CNRS (UMR8617), Universit\'e Paris-Sud, B\^atiment 121, 91405 Orsay, France
\\[-1mm]
%
$^5$ Center for Cosmology and Particle Physics, Department of Physics, New York University, New York, NY 10003, USA
\\[-1mm]
%
$^6$ School of Physics and Astronomy, Cardiff University, The Parade, Cardiff, CF24 3AA, UK 4
\\[-1mm]
%
$^7$ Berkeley Center for Cosmological Physics, University of California, Berkeley, CA 94720, USA
\\[-1mm]
%
$^8$ Lawrence Berkeley National Laboratory, One Cyclotron Road, Berkeley, CA 94720, USA
\\[-1mm]
%
$^9$ Institut de Recherche en Astrophysique et Plan\'etologie, 9, av du Colonel Roche, BP 44346, 31028 Toulouse Cedex 4, France
\\[-1mm]
%
$^{10}$ Jet Propulsion Laboratory, California Institute of Technology, 4800 Oak Grove Drive, Pasadena, CA, USA 91109
\\[-1mm]
%
$^{11}$ Faculty of Electrical Eng., Math. and Computer Sc., Delft Univ. of Technology, Mekelweg 4, 2628 CD Delft,
The Netherlands
\\[-1mm]
%
$^{12}$ SRON, Netherlands Institute for Space Research, Sorbonnelaan 2, 3584 CA Utrecht, The Netherlands
\\[-1mm]
%
$^{13}$ Argelander-Institut f\"ur Astronomie, Universit\"at Bonn, Auf dem H\"ugel 71, D-53121 Bonn, Germany
\\[-1mm]
%
$^{14}$ Department of Astronomy, Cornell University, Ithaca, NY 14853, USA
\\[-1mm]
%
$^{15}$ INAF - Osservatorio di Astrofisica e Scienza dello Spazio, Via Gobetti 101, I-40129 Bologna, Italy
\\[-1mm]
%
$^{16}$ ICC, University of Barcelona, IEEC-UB, Mart\'{\i} i Franques 1, E08028 Barcelona, Spain
\\[-1mm]
%
$^{17}$ Aix Marseille Universit\'e, CNRS, CNES, LAM, Marseille, France
\\[-1mm]
%
$^{18}$ Jodrell Bank Centre for Astrophysics, Dept. of Physics \& Astronomy, The University of Manchester, Manchester M13 9PL, UK
\\[-1mm]
%
$^{19}$ INAF - Istituto di Radioastronomia, Via Piero Gobetti 101, I-40129, Bologna, Italy
\\[-1mm]
%
$^{20}$ INAF - Osservatorio Astronomico di Padova, Vicolo Osservatorio 5, I-35122, Padova, Italy
\\[-1mm]
%
$^{21}$ Institut d'Astrophysique de Paris, 98 bis Boulevard Arago, 75014 Paris, France
\\[-1mm]
%
$^{22}$ Canadian Institute for Theoretical Astrophysics, University of Toronto, Ontario, M5S 3H8, Canada
\\[-1mm]
%
$^{23}$ CAS Key Laboratory for Research in Galaxies and Cosmology,
Dept. of Astronomy, USTC, Hefei, Anhui 230026, China
\\[-1mm]
%
$^{24}$ School of Astronomy and Space Science, USTC, Hefei 230026, China
\\[-1mm]
$^{25}$ Max-Planck-Institut f\"ur Astrophysik, Karl-Schwarzschild Str. 1, 85741 Garching, Germany
\\[-1mm]
%
$^{26}$ Space Research Institute (IKI), Profsoyuznaya 84/32, Moscow 117997, Russia
\\[-1mm]
%
$^{27}$ Instituto de Astrof\'{\i}sica de Canarias, C/ Vía L\'actea 39020 La Laguna (Tenerife), Spain
\\[-1mm]
%
$^{28}$ Physics department, Sapienza University of Rome, Piazzale Aldo Moro 5, 00185, Rome, Italy
\\[-1mm]
%
$^{29}$ INFN sezione di Roma, P.le A. Moro 2, 00815 Roma, Italy
\\[-1mm]
%
$^{30}$ School of Physics, The University of New South Wales, Sydney NSW 2052, Australia
\\[-1mm]
%
$^{31}$ Kavli Institute of NanoScience, Faculty of Applied Sc.,
Delft Univ. Technology, Lorentzweg 1, 2628 CJ Delft,
The Netherlands
\\[-1mm]
%
$^{32}$ Department of Astronomy, University of Maryland, College Park, MD 20742-2421, USA
\\[-1mm]
%
$^{33}$ School of Physics and Astronomy, University of Minnesota / Twin Cities, Minneapolis, MN, 55455, USA
\\[-1mm]
%
$^{34}$ Centro de Estudios de F\'isica del Cosmos de Arag\'on (CEFCA), Plaza San Juan, 1, planta 2, E-44001, Teruel, Spain
\\[-1mm]
%
$^{35}$ Institute for Advanced Study, Princeton, NJ 08540, USA
\\[-1mm]
%
$^{36}$ Center for Computational Astrophysics, Flatiron Institute, 162 5th Avenue, New York, NY 10010, USA
\\[-1mm]
%
$^{37}$ Imperial College London, Blackett Laboratory, Prince Consort Road, London SW7 2AZ, UK
\\[-1mm]
%
$^{38}$ Kavli Institute for Cosmological Physics, University of Chicago, Chicago, IL 60637, USA
\\[-1mm]
%
$^{39}$ Harvard-Smithsonian Center for Astrophysics, 60 Garden Street, Cambridge, MA 02138, USA
\\[-1mm]
%
$^{40}$ NASA/GSFC, Mail Code: 665, Greenbelt, MD 20771, USA
\\[-1mm]
%
$^{41}$ KEK and Sokendai, Tsukuba 305-0801, Japan, and Kavli IPMU, U. of Tokyo, Kashiwa 277-8582, Japan
\\[-1mm]
%
$^{42}$ Department of Physics, Ben-Gurion University, Be'er Sheva 84105, Israel
\\[-1mm]
%
$^{43}$ Institute for Theoretical Particle Physics and Cosmology (TTK), RWTH Aachen University, D-52056 Aachen, Germany
\\[-1mm]
%
$^{44}$ Perimeter Institute for Theoretical Physics, Waterloo, ON N2L 2Y5, Canada
\\[-1mm]
%
$^{45}$ Dipartimento di Fisica e Scienze della Terra, Universit`a degli Studi di Ferrara, Via Saragat 1, I-44100 Ferrara, Italy
\\[-1mm]
%
$^{46}$ INAF - Osservatorio di Astrofisica e Scienza dello Spazio di Bologna, via Gobetti 101, I-40129 Bologna, Italy
\\[-1mm]
%
$^{47}$ Centro de Astrof\'{\i}sica da Universidade do Porto, Rua das Estrelas, 4150-762 Porto, Portugal
\\[-1mm]
%
$^{48}$ Instituto de Astrof\'{\i}sica e Ci\^encias do Espa\c{c}o, CAUP, Rua das Estrelas, 4150-762 Porto, Portugal
\\[-1mm]
%
$^{49}$ Department of Theoretical Physics and Centre for Astroparticle Physics, University of Geneva, 
CH-1211 Geneva, Switzerland
\\[-1mm]
%
$^{50}$ Department of Physics, Harvard University, 17 Oxford St., Cambridge, MA 02138, USA
\\[-1mm]
%
$^{51}$ European Southern Observatory, Karl-Schwarzschild-Strasse 2, Garching D-85748, Germany
\\[-1mm]
%
$^{52}$ Department of Physics \& Department of Astronomy, Yale University, New Haven, CT 06520, USA
\\[-1mm]
%
$^{53}$ Niels Bohr International Academy and Discovery Center, Blegdamsvej 17, 2100 Copenhagen, Denmark
\\[-1mm]
%
$^{54}$ Department of Physics and Astronomy, University of California, Los Angeles, CA 90095, USA
\\[-1mm]
%
$^{55}$ Departamento de Astrof\'{\i}sica, Universidad de La Laguna, E-38206 La Laguna, Tenerife, Spain
\\[-1mm]
%
$^{56}$ California Institute of Technology, 1200 E. California Boulevard, MC 367-17, Pasadena, CA 91125, USA
\\[-1mm]
%
$^{57}$ Department of Physics and Astronomy, University of British Columbia, 6224 Agricultural Road, Vancouver, V6T 1Z1, Canada
\\[-1mm]
%
$^{58}$ Department of Physics and Astronomy, The Johns Hopkins University, 3701 San Martin Drive, Baltimore MD 21218, USA
\\[-1mm]
%
$^{59}$ Department of Physics and Beecroft Inst. for PAC, University of Oxford, 1 Keble Road, Oxford, OX1 3RH, UK
\\[-1mm]
%
$^{60}$ Institute of Theoretical Astrophysics, University of Oslo, P.O.B 1029 Blindern, N-0315 Oslo, Norway
\\[-1mm]
%
$^{61}$ Indian Institute of Science Education and Research (IISER),
Dr. Homi Bhabha Road, Pashan, Pune 411008, India
\\[-1mm]
%
$^{62}$ National Astronomical Observatory of Japan, Mitaka, Tokyo 181-8588, Japan
\\[-1mm]
%
$^{63}$ INFN Sezione di Pisa, Largo B. Pontecorvo 3, 56127 Pisa, Italy
\\[-1mm]
%
$^{64}$ Institute of Astronomy, Madingley Road, Cambridge CB3 0HA, UK
\\[-1mm]
$^{65}$ Kavli Institute for Cosmology Cambridge, Madingley Road, Cambridge CB3 0HA, UK
\\[-1mm]
}

{
\vspace{6mm}
\tableofcontents
\thispagestyle{empty}
}

\setlength{\parindent}{0mm}
\setlength{\parskip}{1mm}

\newpage
\pagenumbering{arabic}
\section*{Executive summary}
\addcontentsline{toc}{section}{Executive summary}
\setcounter{page}{1}

In the past few decades, a standard model of cosmology, $\Lambda$CDM, has emerged.
According to our understanding, our Universe is about 13.8 billion years old, and expanded from an initial hot and dense state. Baryonic matter accounts for 5\% of its present matter-energy content, non-baryonic dark matter for about 25\%, and some unknown form of energy with negative pressure, dubbed dark energy, for the remaining 70\%. The galaxies and clusters of galaxies that we observe today form by the gravitational collapse of tiny primordial fluctuations of the spacetime metric, thought to originate from an early stage of fast accelerated expansion, known as cosmic inflation.

But very fundamental questions remain unanswered: we do not know what the dark matter and dark energy are, whether dark matter interacts, or if extra light particles exist. We are unsure whether inflation did indeed take place and exactly what physics was at work in the very early Universe. We still have to reconcile the laws of gravitation with the standard model of particle interactions -- both of which are known to be incomplete and require extensions to explain existing observations. We do not know the topology of the Universe, or whether it is finite or infinite. We do not fully understand how structure forms, or why some structures on small scales appear to be incompatible with $\Lambda$CDM predictions. We do not have a convincing explanation for anomalies in the large-scale statistics of cosmic microwave background (CMB) anisotropies, except invoking chance multipole alignments and excursions in tails of the realization of a Gaussian random field.

The distribution of matter and energy in the Universe, by virtue of being the product of physical laws and their effect on the Universe's constituents, encodes answers to these questions. We thus propose to conduct an unprecedented full census of this distribution, over scales from one arcminute to the entire sky, and over 99\% of cosmic history. The census will be carried out using a high-angular-resolution, high-sensitivity spectro-polarimetric survey of the microwave sky, which
will track faint signatures of matter and radiation interactions across cosmic time, exploit the CMB as a multi-faceted cosmology probe, and construct a three-dimensional picture of the various components of the cosmic web, across space and time, using five main observables: 
\begin{enumerate}[nosep]
\item tracers of the interaction of the CMB with free electrons in the cosmic web (Sunyaev-Zeldovich effects), to map the distribution of hot gas, its temperature, and large-scale velocity flows;
\item CMB deflections by gravitational lensing, used as a tracer of mass in the entire Hubble volume; 
\item high-redshift dust and line emission, to map atoms in structures across cosmic time;
\item primary CMB anisotropies at the cosmic-variance limit, to constrain parameters of $\Lambda$CDM and its extensions or alternatives;
\item distortions of the CMB blackbody spectrum, to probe the thermal history of the Universe and all processes that can impact it up to redshifts of a few million.
\end{enumerate}

None of these can be observed with high accuracy in full isolation from the others. Most of them probe inter-connected phenomena, which justifies combining the observations. The combined survey will provide a comprehensive and detailed view of the history of the Universe, and a tomographic and dynamic census of the three-dimensional distribution of hot gas, velocity flows, early metals, dust, and mass distribution. In addition to its exceptional capability for cosmology, this survey will be extremely valuable for many other branches of astrophysics.

The proposed survey requires an L-class space mission, featuring three instruments that observe within the 10--2000\,GHz frequency range, at varying spectral and angular resolutions. 
Two of these instruments will be located at the focus of a large (3-m class) cold ($\sim$\,8\,K) telescope, providing arcminute-scale angular resolution at 300\,GHz. A broad-band, multifrequency, polarimetric imager will provide sensitive observations of the CMB and of SZ effects, while a moderate spectral resolution ($R\simeq300$) filter-bank spectrometer will map the IR background and atomic and molecular lines out to high redshift. These instruments will comprise tens of thousands of mm and sub-mm detectors, which will be cooled to sub-kelvin temperatures for sky background-limited performance.

An additional guest focal plane instrument, compatible with the overall constraints of the mission, can be considered for increased science outcomes beyond the mission's primary science goals.

On the same platform (or on an independent spacecraft) absolute spectroscopy across the entire frequency range will be performed by a Fourier-transform spectrometer (FTS) consisting of one or a few independent FTS modules, covering the full 10--2000\,GHz band with spectral resolution ranging from 2.5 to 60\,GHz, angular resolution ranging from a fraction of a degree to a few degrees, and overall sensitivity ${<}\,1\,{\rm Jy}\,{\rm sr}^{-1}$, 4 to 5  orders of magnitude better than that of {\it COBE}-FIRAS.

We envision 6 years of observation from an orbit around the L2 Sun-Earth point, with two different observing modes: a survey for about half the mission time, to map the entire sky as well as a few selected wide fields; and an observatory mode, during which the rest of the time will be made available to the wider scientific community for an opportunity to observe regions of specific interest. 

Although ambitious, the proposed survey builds upon previous space mission concepts already studied at the pre-phase-A or phase-A levels ({\it CORE}, {\it PICO}, and {\it SPICA}). The FTS instrument builds on similar technology flown on {\it COBE\/} and {\it Herschel}, and on the previously proposed {\it PIXIE\/} and {\it PRISTINE\/} mission designs.

Past experience teaches us that progress in cosmology has often come in unexpected ways through opening up new directions.
By systematically probing the Universe using many approaches, and with an unprecedented capability to observe faint signals coming from the largest cosmological distances, the proposed mission will be key in pushing back the frontiers of our understanding of the Universe that we live in. It will be transformational in many areas of physics, astrophysics, and cosmology at the most fundamental level, in a way that is unmatched by any other existing or proposed experiment -- and can only be achieved from space.

While one cannot guarantee completely resolving \emph{all} the open questions we plan to address, the survey proposed is guaranteed to transform our knowledge of the Universe. It will also be of immense legacy value for many branches of astrophysics, with an unprecedented view of microwave emissions at sub-arcminute to few-arcminute angular resolution in hundreds of frequency channels.
\vspace{6mm}
\section{Scientific introduction}

\subsection{Open questions in cosmology}

The twentieth century witnessed the spectacular transformation of physical cosmology into a quantitative branch of science and has ushered the era of ``precision cosmology". 
During the past two decades a standard cosmological model has emerged: inflationary $\Lambda$CDM. Seven independent parameters describe the matter and energy content of the Universe, its expansion history, and the statistical distribution of initial perturbations that evolve to form the large-scale structures we observe today~\citep{2018arXiv180706205P}. 

Even though $\Lambda$CDM provides a good phenomenological fit to most cosmological observations, it lacks an underlying theoretical explanation, and is, at best, strikingly incomplete. In the absence of observations capable of discriminating between different options, parameters in natural extensions to $\Lambda$CDM are set to default values driven by simplicity. For the minimal $\Lambda$CDM, we do not know the nature of 95\% of the contents of the Cosmos. Inflation explains a host of phenomena including the origin of density perturbations and the topology of the Universe, but we do not have a physical model for it.  

Not all current observations match the theoretical expectations of $\Lambda$CDM. For the lowest multipoles, the amplitude of cosmic microwave background (CMB) fluctuations are somewhat too low, enough to motivate investigations for possible explanations. Hemispherical power asymmetry and alignments of CMB multipoles on large scales question the assumptions of homogeneity and statistical isotropy. There is a significant $4\sigma$ tension in the determination of the Hubble constant between CMB and distance-ladder measurements~\cite{2019ApJ...876...85R}, and there are apparent inconsistencies with observations of structures at galaxy and cluster scales~\cite{2017ARA&A..55..343B}.

These inconsistencies may disappear with refinements of theoretical modeling and data analyses, or may point to new physics, as do the existence of dark matter and dark energy, inflation, and the incompleteness of particle physics. 
Are we missing an essential piece of the puzzle? Current data are not sufficient to provide definitive answers.

The matter-energy constituents of the Universe, the laws of gravitation and particle interactions, and the initial state of metric perturbations, all impact the distribution of matter structures and radiation in the Universe. By mapping the components of our Universe \emph{across space and time} at the next level of detail, we will open a new window on the properties of the Universe, and on the machinery that governs cosmic evolution and  encodes the fundamental laws of nature.

\subsection{Cosmological observations in the microwave sky}

Confidence in inflationary $\Lambda$CDM relies on its spectacular consistency with observations of the CMB anisotropy, and with other cosmological probes \citep[Ref.][and references therein]{2018arXiv180706209P}. The primary CMB\footnote{`Primary anisotropy' refers to patterns arising from processes at the last scattering surface'. `Secondary' refers to processes the CMB photons undergo between the last scattering surface and our telescopes.} gives an image of structures and velocity flows in a thin shell at $z\simeq 1100$, and hence also at a specific time. Cosmic variance\footnote{`Cosmic variance' represents variance in measurements arising from the limited statistics of observing a single universe.} precludes stringent statistical tests of the global cosmological paradigm on the largest scales using primary CMB data alone. 

The CMB is a bath of radiation that permeates the entire Universe and interacts with everything in it. Every process that exchanges energy with the cosmic microwave photons leaves an imprint. We propose to exploit these imprints to extract answers to the key open questions in astrophysics by making measurements at microwave and submillimeter wavelengths of four sets of related phenomena:
\begin{enumerate}[nosep]
    \item interactions of the CMB photons with structures of matter through gravitational lensing and through scattering with electrons. Observations of these ``secondary CMB anistropies" will probe structures throughout most of the Hubble volume (Sect.~\ref{sec:backlight});
    \item microwave emission of baryonic matter (cosmic dust, atoms, and molecules) residing in the cosmic web out to high redshift. Mapping this emission and combining it with maps of secondary CMB anisotropies will give a tomographic view of the Universe on large scales (Sect.~\ref{sec:cib-lim});
    \item primary CMB temperature and polarization anisotropies. Higher fidelity measurements will give new insights on the cosmological model and on the fundamental laws of physics (Sect.~\ref{sec:primary});
    \item distortions of the CMB blackbody spectrum. Searching for distortions will open a new window to investigate phenomena predicted by minimal $\Lambda$CDM or its extensions, and will provide unique discovery space for unexpected phenomena (Sect.~\ref{sec:distortions}).
\end{enumerate}

{\it These observables are linked because none of them can be observed in isolation}. All contribute to the total emission observed in the microwave band, and each is a source of confusion when measuring others.
None of these signals can be fully exploited without understanding them all -- and many of them provide answers to closely related cosmological questions.

\section{A census of structures with the CMB as a backlight}
\label{sec:backlight}

Current experiments have vividly demonstrated the enormous potential of using the CMB as a backlight to study the cosmic matter distribution and its evolution.  A high-sensitivity, full-sky survey with arcminute angular resolution and frequency coverage spanning the millimeter and submillimeter wavebands would map the distribution of essentially all baryonic and dark-matter structures in the observable Universe, and measure the peculiar motion of matter within the cosmic web. Such a complete matter census would be transformational.

After leaving the surface of last scattering, CMB photons interact with intervening matter primarily via two processes: (1) scattering by electrons in ionized plasma, called the Sunyaev-Zeldovich (SZ) effect; 
and (2) deflection by gravitational potential wells along the cosmic web. 

There are several variants of the SZ effect. The thermal SZ (tSZ) traces the integral of gas pressure in the baryons associated with large-scale structures along the line of sight. Galaxy clusters are the most dominant contributors. The magnitude of the spectral difference of the tSZ relative to the CMB is called ``the Compton-$y$" parameter. 
The kinetic SZ (kSZ) traces line-of-sight gas momentum with respect to the CMB. 
Relativistic effects (rSZ) subtly alter the tSZ frequency spectrum and can be used to directly measure gas temperatures (e.g., without relying on X-ray data). There are additionally non-thermal SZ (ntSZ) signals that can constrain the particle composition of exotic plasmas, such as radio bubbles driven by AGN feedback, and polarized SZ (pSZ) probes of cluster transverse motions and internal substructure. 

Gravitational potentials lens the CMB backlight \citep{2014A&A...571A..17P,2017PhRvD..95l3529S,2018ApJ...860..137S}. 
By cross-correlating CMB lensing maps with visible tracers, such as galaxies and clusters, we can partition the lensing signal into redshift slices, a process known as ``lensing tomography,'' and measure the growth of structure back to redshifts of a few, well past the point where galaxy cosmic shear becomes ineffective due to lack of background sources. On smaller scales, the lensing effect probes deflections by strong localized over-densities, enabling determination of cluster masses out to redshifts beyond the reach of galaxy shear measurements ($z\,{>}\,2$) \cite{2009arXiv0912.0914L}; this technique will be essential for using high-$z$ clusters as a cosmological tool. CMB lensing can also be used to detect cluster's transverse motions through the moving-lens effect \cite{2019ApJ...873L..23Y, 2018arXiv181203167H}, complementing kinetic and polarised SZ measurements of the cosmic velocity field.

The CMB provides an ideal backlight for these studies because: (1) it originates from a known redshift; (2) its spectrum at emission is known to be an almost perfect blackbody; and (3) its statistical properties are well defined.  It is a new and powerful tool for a comprehensive census of matter in the Universe.

\subsection{Hot gas in the cosmic web}
\vspace{-0mm}

Within only a decade, the number of galaxy clusters detected via the tSZ effect has increased from four to well over two thousand~\cite{2009ApJ...701...32S,2013JCAP...07..008H,2015ApJS..216...27B, 2016A&A...594A..27P}.
A sensitive space mission with close to $1^\prime$ resolution would find {\it all\/} galaxy groups and clusters throughout the observable Universe with masses above $5 \times 10^{13}\,$M$_\odot$, achieving a complete census of these objects from the moment of their appearance. The resulting catalog would yield the ultimate cosmological constraints from cluster counts and clustering, and deliver a goldmine for astrophysical studies of clusters, spawning countless follow-up observations. 

Beyond massive bound objects, the SZ effects open new avenues of research into unbound gas residing in the filaments of the cosmic web in the form of the warm-hot intergalactic medium (IGM), and the circumgalactic medium (CGM) trapped within galaxy dark-matter halos. 90\% or more of all baryonic mass resides in the IGM and CGM.  However, at their characteristic temperatures and densities, these phases defy other means of detection, hiding the majority of baryons under poorly understood conditions.
The SZ effects are an effective tool for mapping these cosmic constituents. A high-precision all-sky map of the Compton-$y$ parameter traces the thermal energy in these phases, while the kSZ effect outlines their spatial density distribution.  Combining them, we can extract key information on the physical state of the IGM and CGM, establishing a whole new class of constraints on viable feedback mechanisms\cite{2017ARA&A..55...59N}. 
More generally, such measurements will finally close our census of the state of all observable matter in the Universe.  These goals cannot be achieved by any other means.

\begin{figure}
\includegraphics[trim=2.2cm 10cm 3cm 3cm,width=0.95\textwidth]{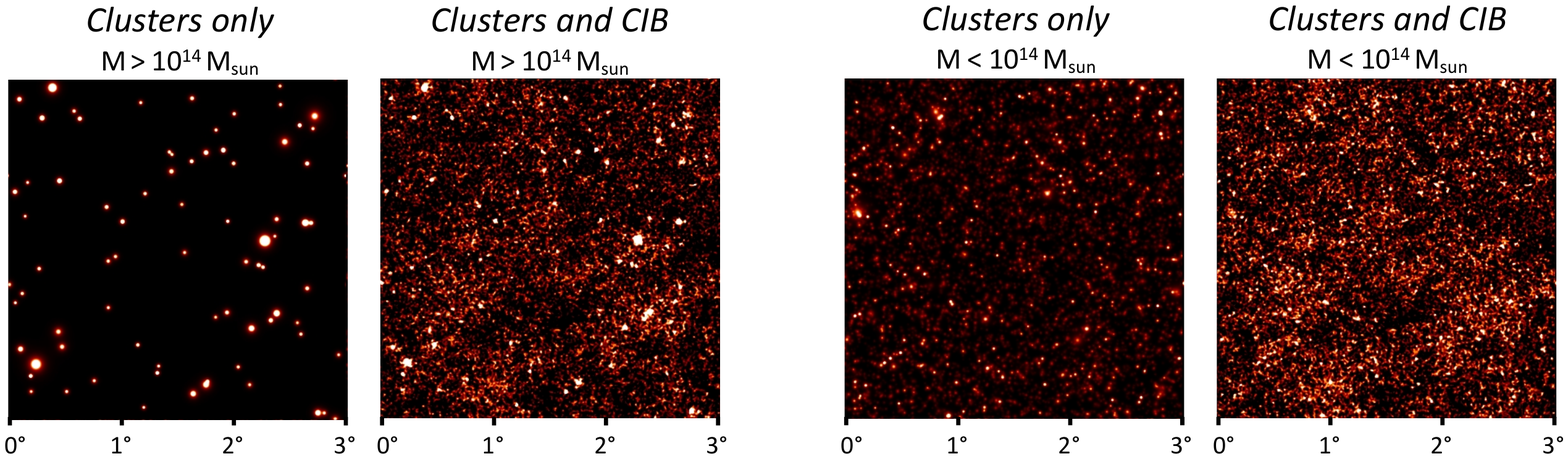}
\vspace{-0.05in}
\caption{ \captiontext Impact of the Cosmic Infared Background (CIB) for detecting low-mass galaxy clusters at 150\,GHz.  {\it Left two panels\/}: simulated clusters of more than $10^{14}{\rm M}_\odot$, without and with 150-GHz CIB fluctuations added. {\it Right two panels\/}: same but with clusters of mass between $1 \times 10^{13}$ and $10^{14}{\rm M}_\odot$. All figures use the same color scale.}
\label{fig:cluster-CIB}
\end{figure}

Removal of astrophysical foregrounds is important at the anticipated signal levels. For example, far-IR thermal emission from dust within galaxies severely contaminates the tSZ signal in lower mass systems (see Fig.~\ref{fig:cluster-CIB}) because the tSZ signal roughly scales as $M^{1.6}$, whereas dust emission is roughly $\propto M$. Observations with multiple frequencies at sub-millimeter wavelengths (from 300\,GHz to 1\,THz) is essential, something that can only be realistically achieved from space.

Wide spectral coverage is also central to the other SZ revolution in the coming decade, namely ``SZ spectroscopy,'' i.e., accurately measuring the SZ spectrum to separate its various components: the rSZ, kSZ, and ntSZ effects (and potentially pSZ too, as well as faint signals from multiple-scattering SZ effects). 
Data from {\it Planck\/} has already motivated studies along these lines~\citep{2019BAAS...51c.302B,Remazeilles2019,Remazeilles2019b}. 

\vspace{1mm}
\subsection{Large-scale velocity flows}

The kinematic SZ effect gives the motion of objects with respect to the CMB rest frame~\cite{1991ApJ...372...21R}.  The cosmic velocity field is a well-known cosmological probe (of dark energy, for example), complementary to measures of the density field.  Furthermore, combining the two makes it possible to test for possible deviations from general relativity on scales that only cosmology can access.  

The kSZ signal is sensitive to the line-of-sight component of the bulk proper motions of ionised gas, and scales with the free-electron optical depth, irrespective of the thermal energy. For clusters and groups of galaxies, the kSZ signal is of the order 1--10$\,\mu$K, just beyond the reach of current-generation CMB experiments for detection in individual objects. However,
cross-correlation with other tracers, particularly galaxy surveys, has produced
statistical detections~\cite{2012PhRvL.109d1101H,2014A&A...561A..97P}.  The scope of kSZ applications will  dramatically expand in the coming decade with wide sky coverage and increasing sensitivity.  This will improve leverage for various cross-correlation studies and also enable individual detections to break the ``optical depth degeneracy'' currently limiting kSZ applications for cosmology. Together with a  smaller contribution from the moving-lens effect via lensing (of the order 0.01--0.1$\,\mu$K), mapping the complete 3D velocity flows in the Universe will be possible.

It is important to note that the spectrum of the kSZ emission law is identical to that of the CMB (ignoring small higher-order effects).
Hence, kSZ signals are best measured on arcminute angular scale in CMB intensity maps free from contamination of foregrounds (including tSZ and the CIB), and with low primary CMB fluctuations.
This underscores the need for high-precision component separation on small scale, which can only be achieved with a large number of frequency channels spanning frequencies that are only observable from space.  

\begin{center}
\begin{figure}[t]
\includegraphics[trim=0cm 6cm 0cm 1cm ,width=\textwidth]{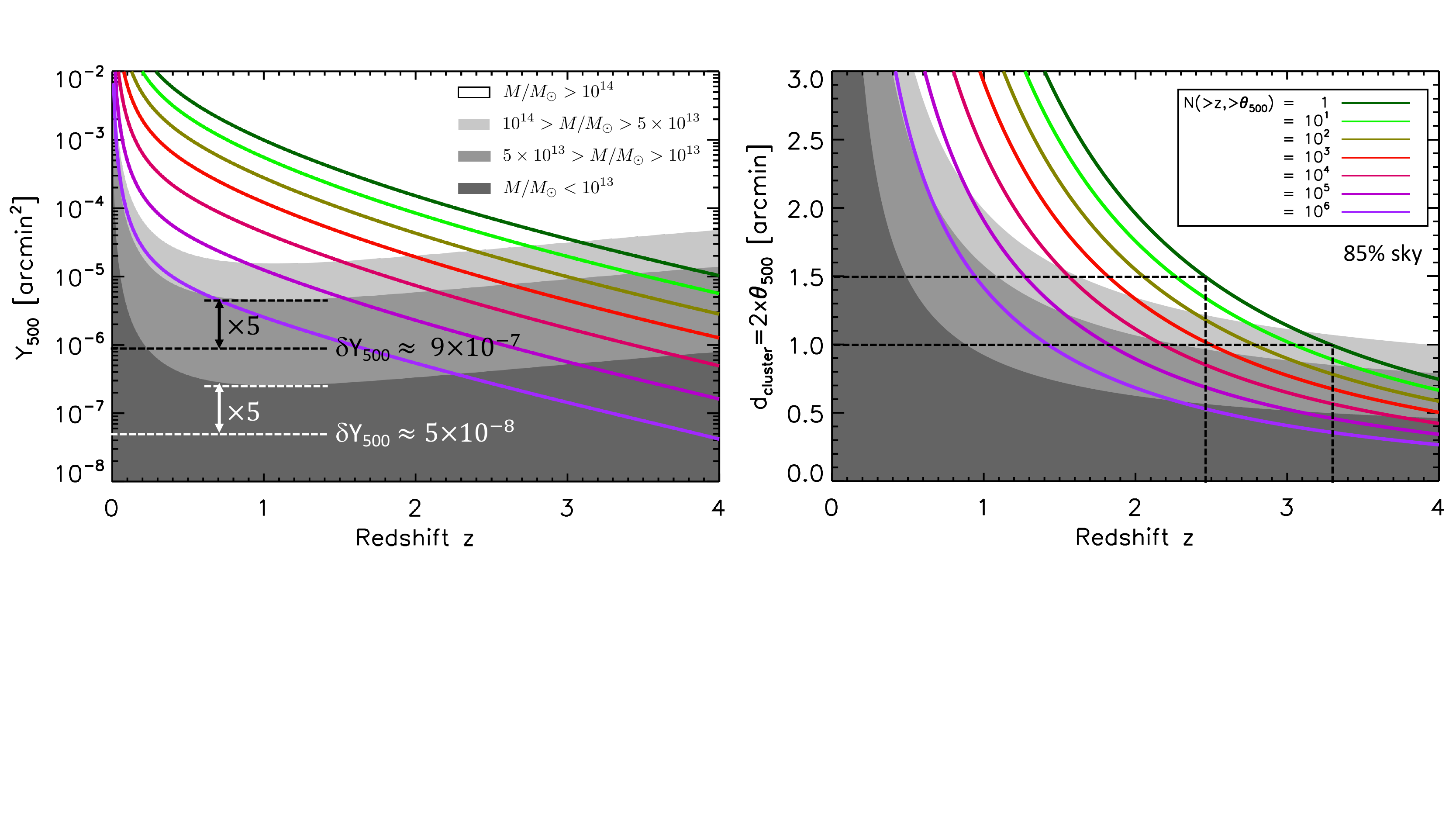}
\vspace{-0.3in}
\caption{ \captiontext {\it Left\/}: Distribution of clusters for various mass ranges as a function of redshift and cluster-integrated Compton parameter $Y_{500}$ (white and gray areas of various shades), modeled following the self-similar model of \cite{2010A&A...517A..92A}. A survey with tSZ flux error $\delta Y_{500} \simeq 9\times 10^{-7}$ would detect \emph{all} clusters of mass
$M>5\times 10^{13}{\rm M}_\odot$ (about 1.5 million objects), while $\delta Y_{500} \simeq 5\times 10^{-8}$ would be sufficient to even detect groups of $\simeq 10^{13}{\rm M}_\odot$. Colored lines show, as a function of $z$ and $Y_{500}$, the expected number of clusters that have both larger tSZ signal, and are located at higher redshift.
{\it Right\/}: Distribution of clusters as a function of redshift and angular size, with the same white and gray color code. All clusters of mass $M> 10^{14}{\rm M}_\odot$ (white) have an angular size larger than $\simeq 1^\prime$, and clusters $>5\times 10^{13}{\rm M}_\odot$ (white and light gray) larger than $\simeq 0.8^\prime$. Colored lines show, as a function of $z$ and cluster angular diameter $d_{\rm cluster}$, the expected number of clusters that have both larger angular diameter, and are located at higher $z$. Dashed lines show that the highest redshift clusters with angular sizes $1.5^\prime$ are at $z \simeq 2.5$, and with size $1^\prime$ at $z \simeq 3.3$.}
\label{fig:clusters}
\vspace{-0.2in}
\end{figure}
\end{center}

\vspace{-1.2cm}

\subsection{Survey requirements for mapping hot gas and velocity flows}
\label{sec:req-sz}

The observation of ionized gas in clusters and filaments that constitute the cosmic web requires the detection of hot gas using the spectral signature of the tSZ effect, and discrimination of the various contributions (i.e., thermal, relativistic, kinematic, and non-thermal effects). 
    
\noindent\myuline{\it Sensitivity and angular resolution:}\quad The left panel of Fig.~\ref{fig:clusters} shows the modeled distribution of clusters of different masses as a function of $Y_{500}$ (integral of the $y$-parameter out to a radius $r_{500}$, the radius inside of which the average density is 500 times the critical density). To detect (at more than 5$\sigma$) all clusters above $5\times 10^{13}{\rm M}_\odot$, we need a sensitivity of $\delta Y_{500}=9\times 10^{-7}$ or better, with the goal of trying to go down to $\delta Y_{500}=5\times 10^{-8}$, to detect groups of $10^{13}{\rm M}_\odot$.

The right panel of Fig.~\ref{fig:clusters} shows that all clusters of mass above $5\times 10^{13}{\rm M}_\odot$ have angular diameter larger than $\simeq 0.8^\prime$. We require an angular resolution on the reconstructed $y$ map of $\simeq 1.5^\prime$. Because of the extended cluster profile, this yields only a maximum sensitivity loss of about 20\% by reason of beam dilution (for detecting an isolated cluster). To reduce blending effects, we target a goal of $1^\prime$ angular resolution.

A cluster with mass $M>5\times 10^{13}{\rm M}_\odot$ has typical Compton optical depth $\tau \simeq 10^{-3}$. For a peculiar velocity of 300\,${\rm km}.{\rm s}^{-1}$, the kSZ amplitude is about 3\,$\mu$K. Detecting this at 5$\sigma$ for a cluster of $1^\prime$ angular size requires a CMB sensitivity of 0.6\,$\mu$K.arcmin (goal). A $3\sigma$ detection can be achieved with a CMB sensitivity of 1\,$\mu$K.arcmin. Measuring kSZ on large clusters is challenging because of confusion with the primary CMB. However, the kSZ effect dominates over the primary CMB at $\ell > 4000$, i.e., angular scales smaller than $3^\prime$. Hence, there is a preference for better angular resolution (requirement $1.5^\prime$, goal $1^\prime$), to measure kSZ for clusters smaller than the primary CMB damping angular scale and hence to avoid much of the extra CMB noise (see left panel of Fig.~\ref{fig:lensing-req}). 

\noindent\myuline{\it Frequency range and number of channels:} \quad The survey must allow one to separate the tSZ and kSZ signals of interest from other astrophysical sources of emission, in particular the small-scale thermal dust and radio emission from extragalactic sources, either those associated with the cluster, or those that constitute the overall radio and infrared source background and contribute to the overall sky noise.

In fields away from Galactic contamination, using at least 3 frequency channels to detect and characterize radio-sources, at least 3 for IR sources, and around 6 for the various SZ effects and the primary CMB with some redundancy, leads to at least a dozen frequency channels being required for separating the various components when all of those are above the instrumental noise (as is likely to be the case for many of the interesting clusters detected by a sensitive survey). Over a large fraction of sky, extra channels are needed also to isolate small-scale emission from the Galactic interstellar medium (i.e., synchrotron and free-free, plus thermal and anomalous dust emission), for a total of around 20 frequency bands. These channels should cover the frequency range where the SZ signals are the strongest relative to other emissions (between 100 and 400\,GHz), with channels around the tSZ minimum, null, and maximum (150, 220, and 350\,GHz), and extra channels at lower and higher frequencies to characterize low- and high-frequency foregrounds. 

As seen in Fig.~\ref{fig:cluster-CIB}, fluctuations of the CIB are a serious source of noise for detecting clusters below $10^{14}{\rm M}_\odot$, even at 150\,GHz; however, if the CIB can be reduced to $\simeq 20$\% of its initial amplitude, tSZ emission from clusters below $10^{14}{\rm M}_\odot$ become detectable. This should be feasible over most of the sky if the survey includes observations between 300 and 800\,GHz, where CIB fluctuations dominate on small scales. Such observations must be done from space.
CIB fluctuations are still more problematic for measuring the fainter rSZ correction (typically less than 10\% of the tSZ, and seen at higher frequency). A wide spectroscopic survey of the CIB in the 300--1000\,GHz frequency range, allowing for decomposition into contributions from different redshifts and/or components, will help characterize and subtract this contaminant to the best possible accuracy.  
\noindent
\begin{center}
\fbox{\begin{minipage}[h][][t]{0.98\textwidth} 
\textbf{Requirements and goals for mapping ionized gas distribution and velocity flows}

Thermal SZ sensitivity at 150 and 350\,GHz of $\sigma_y \simeq 10^{-6}$ per $\simeq 1^\prime $ pixel, or better;\\
Kinematic SZ sensitivity of $\Delta T \simeq 1\,\mu$K (requirement) to $0.6 \, \mu$K (goal) per $\simeq 1^\prime $ pixel;\\
Angular resolution $1.5^\prime $ or better ($1^\prime $ goal);\\
Full-sky observations from 50 to 800\,GHz with $\simeq $\,20 frequency channels;\\
Spectroscopic observations of the CIB in the 300--1000\,GHz frequency range.
\end{minipage}}
\end{center}

\subsection{A survey of matter through CMB lensing}

\begin{center}
\begin{figure}
\includegraphics[trim=2.5cm 6cm 3cm 3cm ,width=\textwidth]{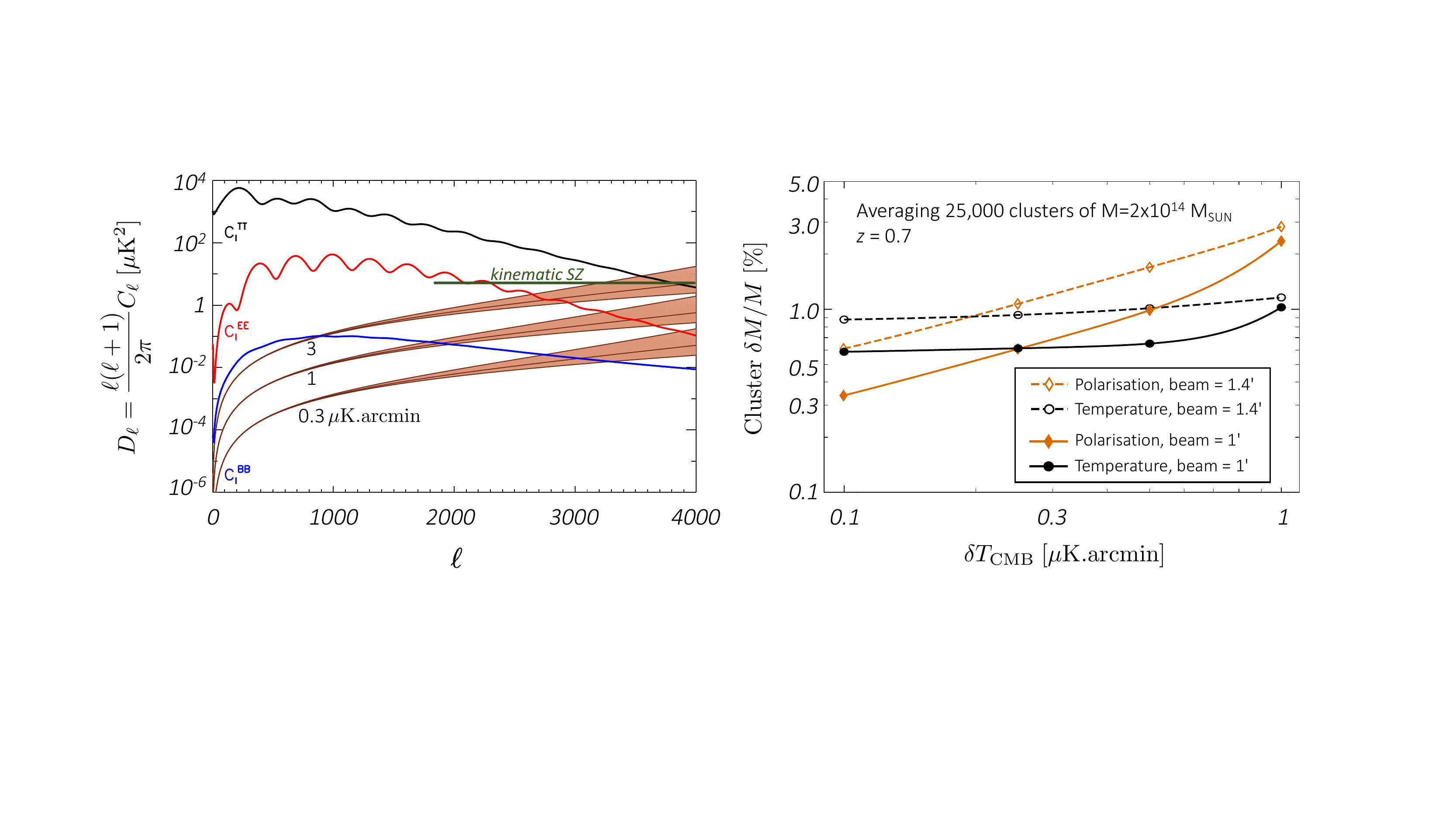}
\vspace{-0.2in}
\caption{\captiontext {\it Left\/}: CMB $TT$, $EE$, and lensing $BB$ spectra. The light brown bands correspond to noise at the level of 3, 1, and 0.3\,$\mu$K.arcmin, and angular resolution ranging from 1 to 3 arcmin. The dark green horizontal line shows the approximate level of the kSZ effect on small scales ($\ell>2000$, \cite{2013JCAP...07..025D}). {\it Right\/}: Accuracy of cluster mass calibration achieved by averaging 25,000 clusters at redshift 0.7, both from temperature and from polarization measurements, for resolutions of $1.4^\prime$ and $1^\prime$.}
\label{fig:lensing-req}
\vspace{-0.2in}
\end{figure}
\end{center}

The lensing of the CMB by gravitational potentials along the line of sight is of considerable interest in two main regimes. First, very large-scale structures, for which the growth is still linear, generate deflections of the path of the photons with typical amplitude of $3^\prime$. These deflections are coherent over scales of few degrees. Maps of the deflections can be used to reconstruct maps of the gravitational potential on the largest scales. The fidelity of such reconstruction is determined  by the sensitivity to the lensing $B$ modes.\footnote{The angular power spectra of the polarization of the CMB are encoded in terms of $E$ and $B$ modes. Inflation, density perturbations, and foregrounds produce $E$ modes. Inflation, lensing, and foregrounds each produce distinct pattern of $B$ modes.} If a survey's $B$-mode sensitivity is better than the level of the $B$-mode lensing signal ($5\,\mu$K.arcmin amplitude), signal-dominated maps of the lensing potential can be reconstructed, down to $20^\prime$ for a {\it CORE}-like CMB polarization survey \citep{2018JCAP...04..018C}, and about $10^\prime$ or better with higher sensitivity, as in the case of {\it PICO\/} \citep{2019arXiv190210541HB}.
Secondly, on much smaller scales, dense, collapsed objects, such as massive clusters of galaxies, also deflect the CMB backlight. This can be used to estimate the mass of the lensing object, as illustrated on the right panel of Fig.~\ref{fig:lensing-req}.

\vspace{-0.3in}

\subsection{Survey requirements for CMB lensing}

The reconstruction of maps of integrated lensing potential and the calibration of the cluster masses require high-sensitivity observations of the CMB temperature and polarization, in particular on small scales. Lensing reconstruction can be performed with various combinations of CMB maps. The $TT$ estimator uses intensity only, while the $EB$ estimator, which is the most effective for very low noise, uses $EB$ cross-correlation \citep{2002ApJ...574..566H}.

\noindent\myuline{\it Sensitivity and angular resolution:}\quad It has been shown that in the case of a survey with 3$\,\mu$K.arcmin CMB intensity sensitivity, the quality of the lensing reconstruction saturates for an angular resolution of 2--$4^\prime$, while for a $4^\prime$ angular resolution survey, it saturates for a map sensitivity of the order of 0.1--0.3$\,\mu$K.arcmin (see figure~4 of \citep{2002ApJ...574..566H}). 

Figure~\ref{fig:lensing-req} illustrates how varying the sensitivity and angular resolution of a survey changes the performance of lensing measurements. As shown in the left panel, for a noise level of $3\,\mu$K.arcmin, lensing $B$ modes are dominant over the noise up to a limiting $\ell_{\rm lim} \simeq 800$, independently of the angular resolution (here from 1$^\prime$ to 3$^\prime$). Hence, the number of harmonic modes that can contribute to lensing reconstruction is independent of the angular resolution for this instrumental sensitivity. When the sensitivity is increased to 1 or $0.3\,\mu$K.arcmin, $\ell_{\rm lim}$ grows substantially, increasing the number of modes useful for lensing reconstruction with the EB estimator by a factor 4--10 (for lower noise, $\ell_{\rm lim}$ becomes dependent on the angular resolution).

At the same time, we note that the kSZ effect becomes the dominant source of confusion for lensing reconstruction with $TT$ when the noise level is below $3\,\mu$K.arcmin. This confirms the sensitivity requirement to measure the kSZ effect (Sect.~\ref{sec:req-sz}), and shows that only polarization can improve lensing reconstruction when the performance of the instrument exceeds this level.

The right panel shows a confirmation from estimates of the accuracy of the calibration of cluster mass from stacking lensing constraints for a population of 25,000 clusters at $z=0.7$. The sensitivity saturates for a low noise level in $TT$, while the polarization constraints keep improving. We see from the right panel of Fig.~\ref{fig:lensing-req} that if the sensitivity is at the level of about 1$\,\mu$K.arcmin, an angular resolution of $1.4^\prime$ or better allows for calibrating cluster masses at about the 1\% level with $TT$ only, which outperforms polarization constraints. With a sensitivity 0.6$\,\mu$K.arcmin and a $1^\prime$ beam, we improve the constraint by a factor of 2. To obtain comparable constraints with polarization (as a cross-check, and for an improvement by $\sqrt{2}$ of the overall sensitivity), we need a sensitivity better than 0.3$\,\mu$K.arcmin, and a $1^\prime$ beam.

\noindent\myuline{\it Frequency range and number of channels:}\quad Requirements in terms of frequency range and frequency channels are similar for lensing as for measuring SZ effects (the kSZ effect has the same electromagnetic spectrum as the lensing signal).
\noindent
\begin{center}
\fbox{\begin{minipage}[h][][t]{0.98\textwidth} 
\textbf{Requirements and goals for mapping CMB lensing to trace the distribution of mass}

Full-sky CMB sensitivity $\Delta T$ per $\simeq 1^\prime$ pixel between $\simeq 1\,\mu $K  (requirement) and $\simeq 0.6\,\mu $K (goal);\\
Deep-patch CMB sensitivity $\Delta T$ per $\simeq 1^\prime$ pixel of $\simeq 0.3\,\mu$K;\\
Angular resolution between $1.4^\prime $ (requirement) and $1^\prime $ (goal);\\
Observations from $\simeq50$ to $\simeq 800\,$GHz with $\simeq$\,20 frequency channels.
\end{minipage}}
\end{center}

\section{High-redshift structures on the largest scales}
\label{sec:cib-lim}

While the first stars reionize the Universe at redshift $z \simeq 8$, they also convert a fraction of the primordial hydrogen and helium into heavier atoms (metals), which then form molecules and dust particles that emit radiation at (sub)mm wavelength through thermal and line emission. Their detection out to high redshift, across large patches of sky, in hundreds of frequency bands, opens the path to a census of baryons in these various forms across cosmic time, and hence to the star formation history. Fluctuations of this emission traces the cosmic web at high redshift, and hence structures in a large fraction of the Hubble volume.

\subsection{Revealing galaxy protoclusters via dusty starbursts}

Understanding the full evolutionary history of
galaxy clusters \citep{kra12}, the largest virialized structures in the Universe, is of fundamental importance for the observational validation of the formation history of the most massive dark-matter halos, a crucial test of models for structure formation, as well as for investigating the impact of environment on the formation and evolution of galaxies. Because of their deep potential wells, clusters may preserve fingerprints of the physical processes responsible for triggering and suppression of star formation and black-hole activity. Historically, clusters of galaxies have also been powerful probes of cosmological parameters.

Galaxy clusters in formation are called {\it protoclusters} \citep{ove16}. Above $z \simeq 2$ protoclusters are found to be bright at far-IR/sub-mm wavelengths \citep{dan14,cle16} because a substantial fraction of their member galaxies, the dusty star-forming galaxy population (DSFGs, \cite{cas14}), are undergoing intense, dust-enshrouded star-formation activity. They are rich in molecular gas and heavily obscured by dust. They are thus prime targets for sub-mm observational facilities.

\begin{figure}
\includegraphics[width=1.00\textwidth]{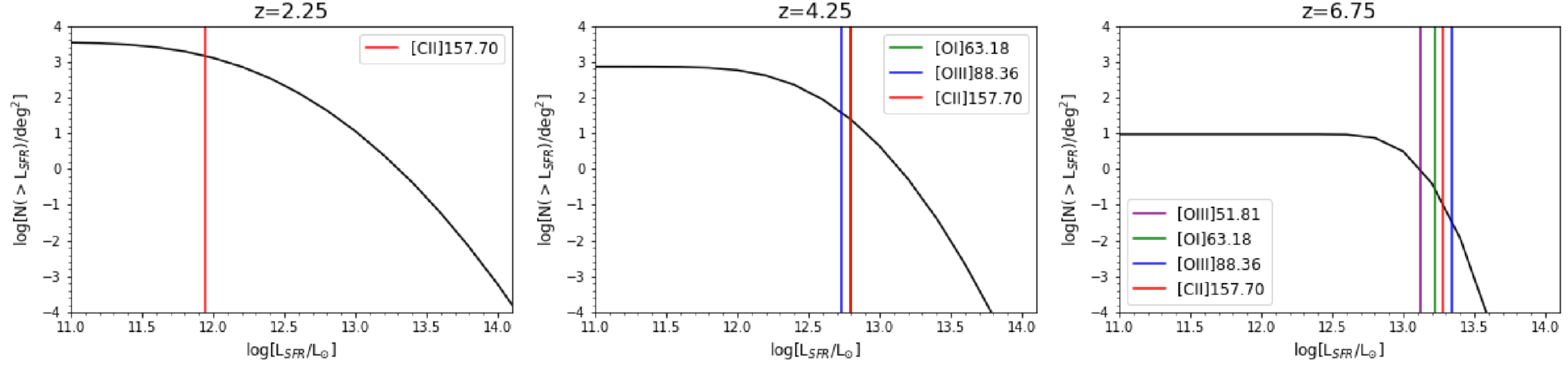}
\vspace{-0.25in}
\caption{\captiontext Cumulative IR (8--$1000\,\mu$m) luminosity functions within $\delta z= 0.5$ at three redshifts. The predictions are based on the model by \citet{neg17}.  The line luminosities corresponding to $L_{\rm IR}$ were computed as described in the text. The vertical lines show the detection limits for the brightest lines, assuming the instrument-performances quantities described in the text. Such an instrument will detect protoclusters of dusty galaxies all the way out to the re-ionization redshift. }
\label{fig:protospheroids}
\vspace{-0.2in}
\end{figure}

Consider a 2-yr full-sky spectro-imaging survey with a 3.5-m telescope and a 64-pixel filter-bank camera observing from 150 to 750\,GHz with $R=300$, an optical efficiency of 30\%, and instrumental noise close to the photon noise. Such an instrument has FWHM ranging from $2.5^\prime$ at 150\,GHz to  $0.5^\prime$ at 750\,GHz and a $5\,\sigma$ point source detection limit ranging from $5\,$mJy at 750\,GHz to $28\,$mJy at 150\,GHz. The corresponding line detection limits are $5\times 10^{-20}$ and $1.4\times 10^{-19}\,\hbox{W}\,\hbox{m}^{-2}$, respectively.  

Such a spectroscopic survey can go deeper than the broadband \textit{Herschel} surveys, with a similar telescope size, because they can take full advantage of the extreme sensitivity of state-of-the art instruments, and do not have the same confusion-noise limit as continuum surveys.

With the resolution of the example instrument, protocluster cores, having sizes of a few 100 kpc \citep{Ivison13,Wang2016,Miller2018,Oteo2018}, are unresolved clumps of DSFGs and so will show up as extremely bright sub-mm sources. We expect the detection of the strongest sub-mm lines for thousands of them all the way to the reionization epoch (see Fig.~\ref{fig:protospheroids}). At the peak of cosmic star-formation activity ($z=2$--3) we expect the detection of hundred of thousands protoclusters. At $z>2$ at least two lines will be detected, allowing a solid redshift determination without requiring follow-up observations.

No other foreseen survey can do anything similar. For example, {\it Euclid\/} will detect galaxy protoclusters up to $z\simeq2.5$ and will miss those with strong obscuration by dust. The dynamics and chemistry
of the discovered protoclusters can be further studied with ground-based large-aperture sub-mm telescopes and interferometers including the Atacama Large-Aperture sub-mm/mm Telescope (AtLAST) \citep{dan19b,kla19}, NOEMA and ALMA, underscoring the survey's complementary with these efforts.

As a by-product of our systematic search for protoclusters, we expect to identify tens of thousands of strongly lensed dusty starbursts \citep{DeZotti2019} out to $z\simeq 7$ or even higher.  Preliminary studies of this kind have been carried out for {\it Planck}-selected sources \citep{can15,har16,dan19a,fry19}.  High-resolution follow up (aided by strong lensing) of such exceptionally bright galaxies will provide direct information on the complex physics that governs galaxy formation and early evolution.

\subsection{Tomography of the cosmic infrared background}

The CIB is a major component of the extragalactic background light, 
with a spectrum that spans wavelengths from the millimeter regime down to the mid-infrared \cite{2005ARA&A..43..727L,2018ApSpe..72..663H}. The CIB is sourced by emission from starlight-heated dust in galaxies with a young population of stars. Moderate resolution spectroscopy ($R \simeq 100$) with wide frequency coverage can yield tomographic CIB maps that trace galaxy evolution across time. 

Measurements of both the mean intensity and the spatial fluctuations of the CIB can yield important scientific output. Reaching high accuracy in absolute measurements is particularly challenging. Current constraints from the combination of absolute photometry from FIRAS with relative photometry from \textit{Planck} are only at the $10\%$ level~\cite{Odegard:2019drh}. A future space mission with an absolute spectrometer will allow more than an order-of-magnitude improvement in precision and enable the detection of extended intergalactic dust emission (or emission from more exotic sources such as dark-matter decay~\cite{Creque-Sarbinowski:2018ebl}), by separating out the signal coming from known galaxy populations. 

Meanwhile, measuring the anisotropies in the CIB can be used to probe properties of the host halos of DSFGs \cite{Maniyar:2018xfk}. However, this approach is only useful up to $z\simeq3$ due to large degeneracies between the effects caused by different sources of the anisotropy \cite{Bethermin:2013nza}. Therefore, at higher redshift it is more useful to use cross-correlation with other tracers of galaxy clustering and star-formation history, so that the redshift information is used to break the degeneracies. The CIB fluctuations signal can be fully exploited in cross-correlation with either spectroscopic galaxy surveys at low redshifts or line-intensity maps (e.g., of CO and {\sc Cii} emission) at medium and high redshifts. Cross-correlation with CMB lensing is also promising, particularly near the peak of the CMB lensing kernel, which roughly coincides with the peak of the star-formation rate. CIB fluctuations can also provide an increase in the number of modes used in cosmological analyses at high redshift, improving the constraining power on primordial non-Gaussianity~\cite{Tucci:2016hng}, for example.

\subsection{Mapping first stars and first metals}

An instrument targeting a wide range of frequencies with good spectral resolution can map the intensity fluctuations in multiple molecular and atomic lines across a correspondingly wide range of source redshifts.
Line-intensity mapping (LIM) \cite{Kovetz:2017agg} -- a measurement of spatial fluctuations in the integrated spectral-line emission  originating from many individually unresolved galaxies and from the diffuse intergalactic medium -- makes it possible to track the growth and evolution of cosmic structure at otherwise inaccessible redshifts. 

\begin{figure}[th]
\centering
\includegraphics[width=0.9\linewidth]{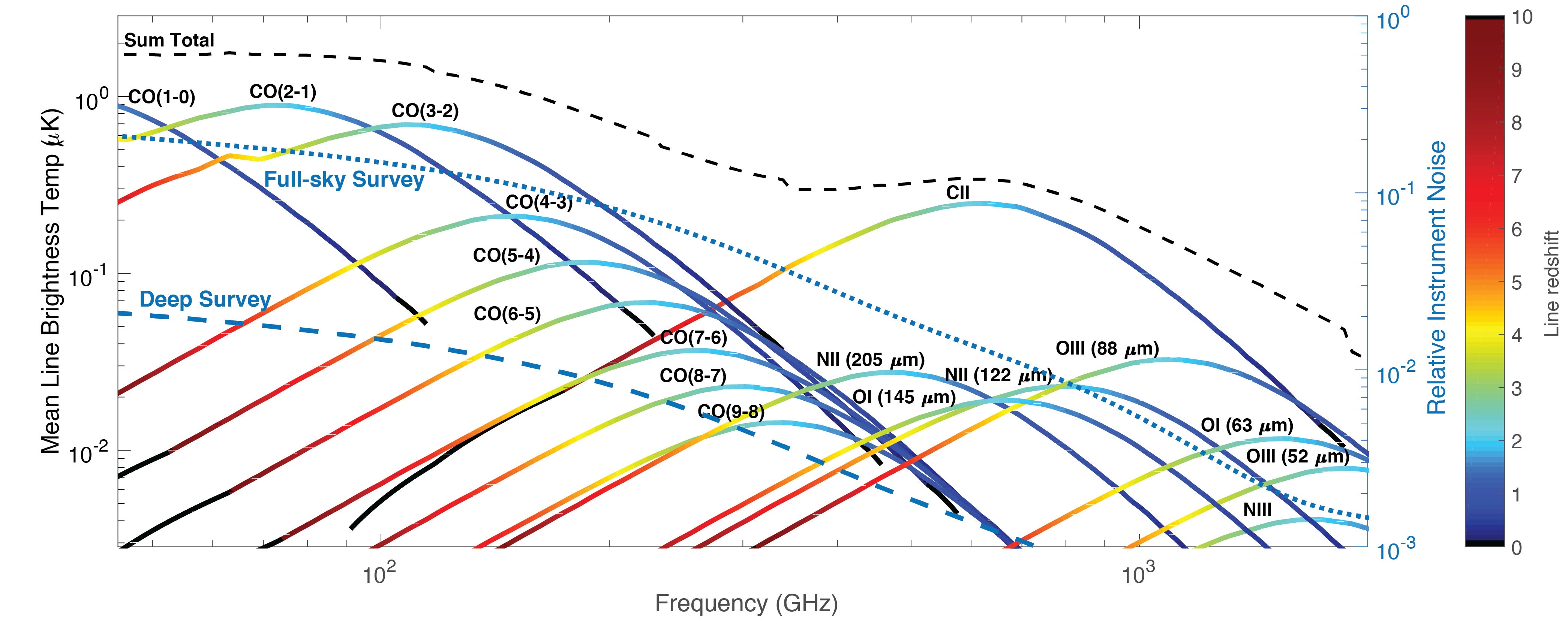}
\vspace{-4mm}
\caption{\captiontext Line emission from various extragalactic atoms and molecules. The sensitivity per ${\rm deg}^2$ expected from a 2-year full-sky survey (dotted blue line) and a deep intensity-mapping survey (dashed blue line) is also plotted. We assume 64 spectrometric pixels with $R=300$ at the focus of an 8-K, 3.5-m telescope, at the photon noise limit, with efficiency 30\%. The full-sky survey is expected to detect [{\sc Cii}] emission from dust heated by star formation with high S/N out to $z\simeq5$, and [{\sc Oiii}], [{\sc Nii}] at lower redshift. A deep survey can measure the CO ladder out to a redshift of 4. The line strengths were modelled as a function of IR luminosity using observationally-based scaling relations \citep{2012ApJ...747L..31S,Bonato2019}. These measurements include local, and high-z galaxies; due to the lack of better constraints, the line ratios are assumed to be constant with redshift. The IR luminosities used to compute the line intensities, and bias was derived from the SFR in the Eagle (Evolution and Assembly of GaLaxies and their Environments) simulation \citep{2015MNRAS.446..521S}, which constraints these relations from as a function of redshift. The SFRD derived from the Eagle simulation shows a good fit to UV based observational constraints. Due to the lack of observational constraints, the modelling of these lines is uncertain by a factor of a few (low redshift)  to an order of magnitude towards high redshift ($z>6$). The modelling of these lines with other analytical calculations and galaxy simulations face similar problems due to a lack of constraints on a large number of free parameters \citep[see e.g.][]{2015JCAP...11..028M}. These line models can only be meaningfully improved with a LIM space mission like the one proposed here. }
\label{fig:IMVisual}
\vspace{-0mm}
\end{figure}

Figure~\ref{fig:IMVisual} shows targets for line-intensity mapping in the frequency range of interest. Notably, these include higher rotational transitions of the carbon-monoxide (CO) molecule, and the $158\,\mu$m {\sc Cii} fine-structure line. The former provide excellent tracers of molecular gas evolution in galaxies, while the latter provides strong constraints on dust-obscured star formation, and is a key probe of galaxies during the epoch of reionization. Additional accessible lines include [{\sc Nii}] and [{\sc Oiii}] from stellar/{\sc Hii} regions, plus [{\sc Oi}] and [{\sc Ci}] from photo-dissociation regions (PDRs), among others.

Detection of multiple lines from the same structures is especially useful. First, this would allow a proper separation of the different line emission components contained in the intensity maps, which is essential to making robust interpretations of the measurements. Secondly, information from different lines can enhance the scientific output, e.g., probing the distribution of gas densities and temperatures in molecular clouds hosting star formation. At low $z$ the CO ladder can be used to constrain molecular gas with a much higher precision than with just one line, while at high $z$ the overall CO emission in high-$J$ transition lines can probe important properties such as gas turbulence, star-formation efficiency, metallicity, and the strength of the ionizing radiation field.  Synergies between different lines can be used to characterize the physical processes that govern reionization, which span various scales and different regions within the first galaxies. In addition, the large-scale fluctuations in this emission should correlate with the overall morphology of the reionization field, and LIM of metal lines from the first galaxies will shed light on the timeline for metal and dust enrichment of the Intergalactic medium (IGM).   

\begin{figure}
\centering
\vspace{-5mm}
\includegraphics[width=0.9\textwidth]{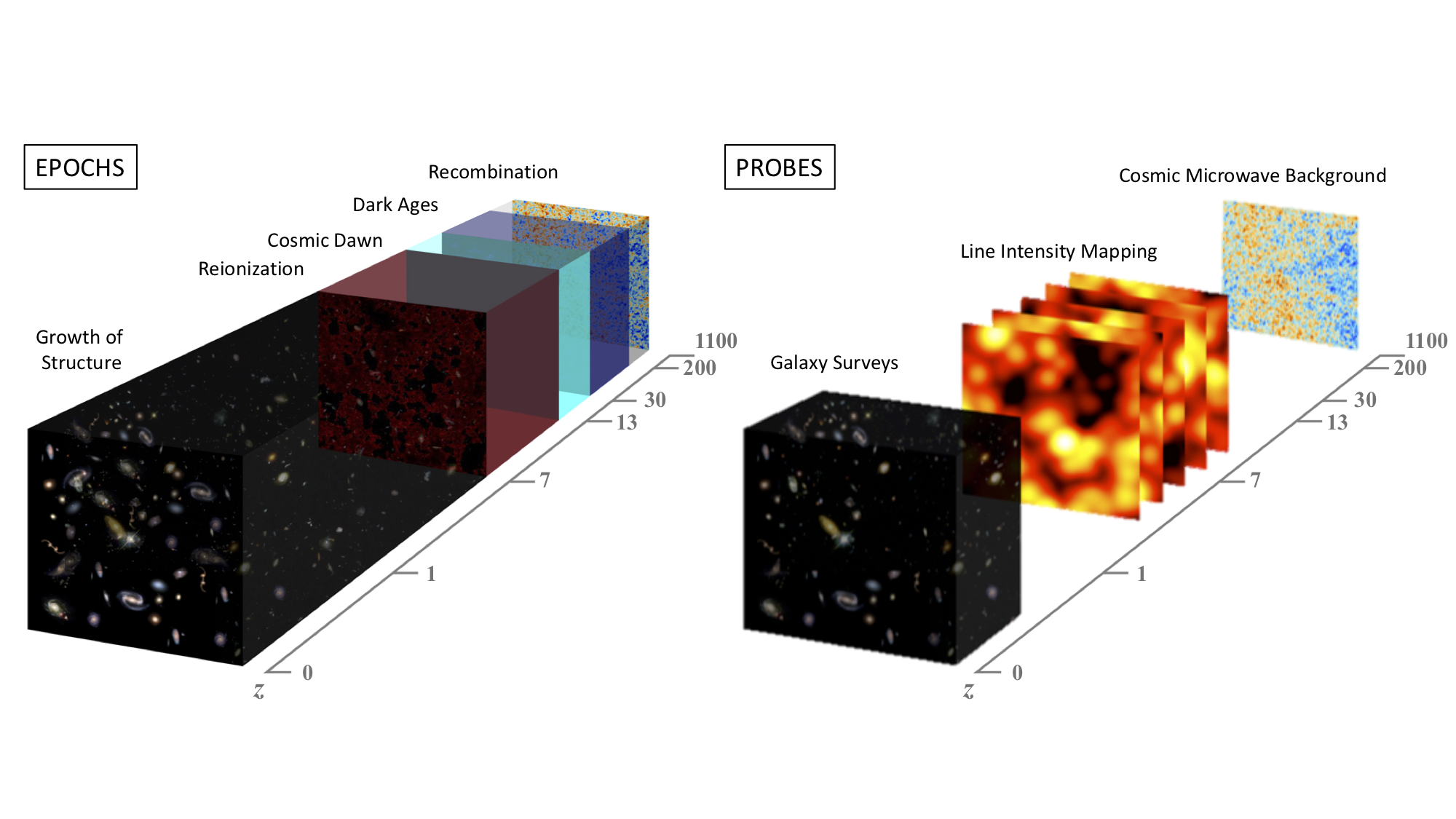}
\vspace{-10mm}
\caption{\captiontext {\it Left\/}: Various epochs in cosmic history, from CMB emission at the time of recombination, to present times. {\it Right\/}: CMB anisotropies map a shell of the Hubble volume located at $z \simeq 1100$. Line-intensity mapping (and CIB tomography) map the large-scale distribution of matter over a wide range of redshifts that cannot be easily accessed by any other means and are at higher $z$ than typical galaxy surveys.}
\label{fig:lim}
\vspace{-0.15in}
\end{figure}

\subsection{Perspectives for cosmology with line-intensity mapping}

Until now, constraints on the parameters of $\Lambda$CDM have come from two main sources, the CMB and galaxy surveys, which originate from high and low redshifts. Over the wide intervening redshift range, between the last-scattering surface and the reach of galaxy surveys, we currently lack observables that provide low uncertainty measurements (Fig.~\ref{fig:lim}). Measurements with LIM across this uncharted volume have the potential to increase cosmological parameter constraining power, and they hold unique qualitative advantages over the more established observables. For example, non-linear effects come in at smaller scales as we go to higher redshifts, allowing more robust comparison to theoretical calculations. 
Ultimately, the combination of  experiments targeting  different  observables will be most effective in breaking different degeneracies between the cosmological parameters.

As well as improving $\Lambda$CDM constraints, LIM has the potential to shed light on the nature of dark matter, dark energy, and what drives early-Universe inflation. Intensity mapping of CO rotational lines at medium redshifts and {\sc Cii} emission towards reionization can fill the gap in measurements of the cosmic expansion history~\cite{Bernal:2019gfq}, which may be crucial for understanding the growing tension with measurements of the Hubble constant and whether it involves time-dependent dark energy. The large number of modes can yield constraints on primordial non-Gaussianity at the level of $f_{\rm NL}\lesssim1$~\cite{MoradinezhadDizgah:2018lac}, which is the target threshold for discerning between single and multi-field models of inflation. Mono-energetic dark matter decay can be tested using its correlation with the mass distribution inferred from the cross-correlation of spectral-intensity maps with galaxy or weak-lensing surveys~\cite{Creque-Sarbinowski:2018ebl}. And probing beyond $\Lambda$CDM, powerful constraints can be placed on the number of effective relativistic degrees of freedom and the sum of neutrino masses~\cite{Bernal:2019jdo}.

\subsection{Survey requirements for high-redshift science}

The study of the high-redshift universe as described above combines the detection of three types of emissions: i) diffuse dust emission from the background of unresolved galaxies constituting the CIB; ii) emission from unresolved galaxies with LIM of metal lines for mapping large scale structure (LSS) and its cosmic evolution; iii) emission from compact sources (continuum and lines) at high redshift.

Mapping diffuse emissions can be done with moderate angular resolution ($\sim 5^\prime$). To map CII at $z>5$, up to the epoch of reionization, we need a sensitivity of 0.03\,$\mu$K at 200 to 400\,GHz (Figure~\ref{fig:IMVisual}). The four first lines of the CO ladder can be mapped between 50 and 200 GHz with sensitivity 0.2\,$\mu$K. We require detecting at least two of them at each redshift bin together with {\sc Cii} so that we can separate the different lines.

As the level of these emissions is uncertain, we require that the space mission have the capability, in addition to the full sky survey, to map a deep patch of sky of a few hundred square degrees.

Detecting individual high redshift objects requires the best possible angular resolution. However, with LIM, even with a resolution of a few arcminutes a survey matching the above sensitivity requirements would detect many high redshift protoclusters and strongly lensed dusty galaxies.

\noindent
\begin{center}
\fbox{\begin{minipage}[h][][t]{0.98\textwidth} 
\textbf{Requirements and goals for high-z science with (sub-)mm dust and line emission}

Sensitivity per $\simeq 1^\circ$ (or smaller) pixel better than 0.03\,$\mu$K in the 200--400\,GHz frequency range, and better than 0.1\,$\mu$K in the 50--200\,GHz frequency range;\\
Angular resolution between $5^\prime $ (requirement) and $1^\prime $ (goal);\\
Spectroscopic observations from $100$ to $1000\,$GHz with spectral resolution $R \simeq 300$;\\
Capacity to observe deep patches.
\end{minipage}}
\end{center}

\section{Cosmology and fundamental physics}
\label{sec:primary}

\begin{wrapfigure}{R}{0.45\textwidth}
\centering
\includegraphics[width=0.5\textwidth]{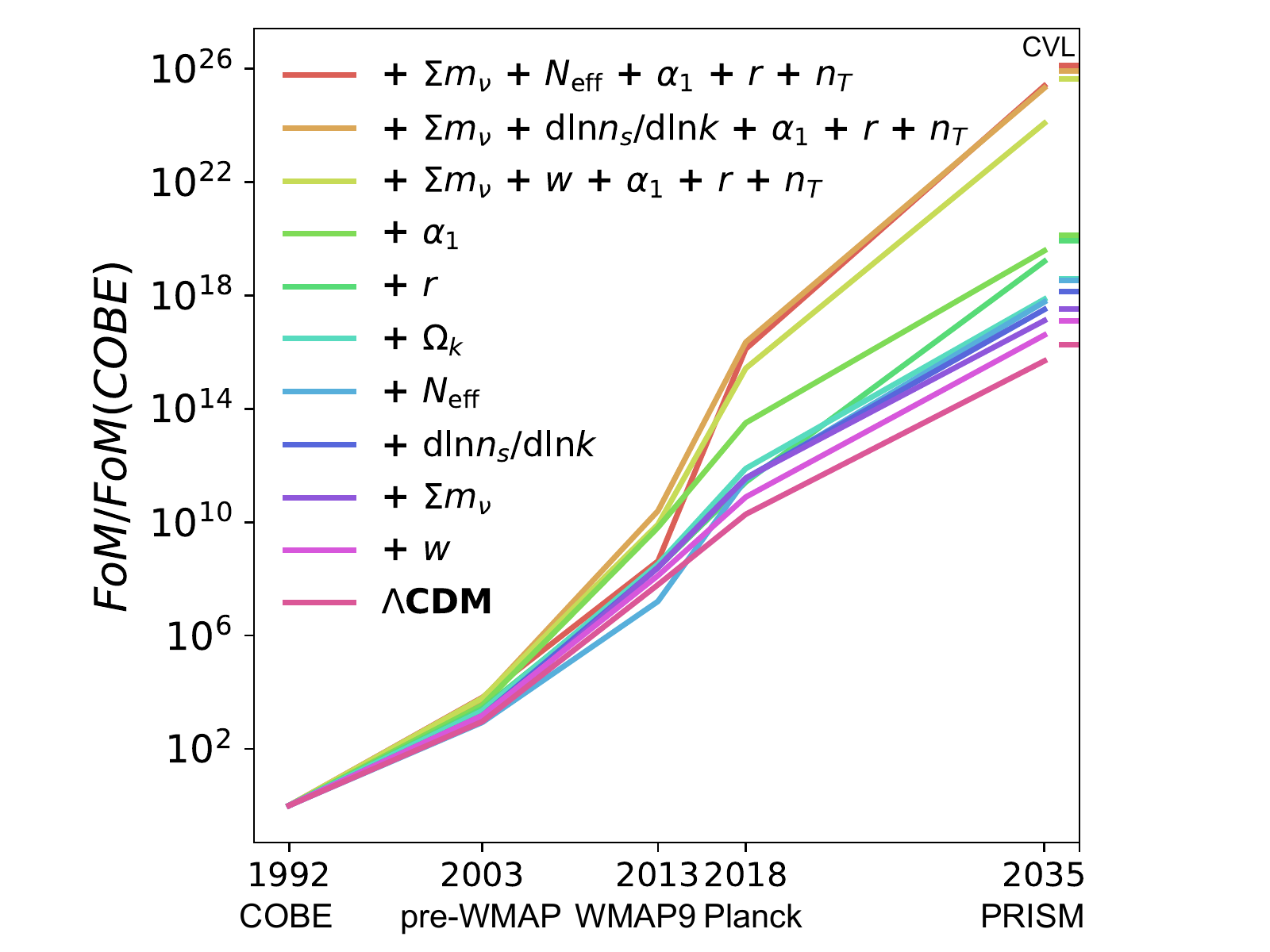}
\caption{\captiontext FoM improvement since COBE for $\Lambda$CDM and several extended cosmological models. For {\it PRISM\/} in 2035, we consider an instrument with {\it PICO\/}-like channels and sensitivity, and 2.5 times smaller beams. The constraints on $\Lambda$CDM and extensions reach the cosmic variance limit (CVL), shown as horizontal dashes on the right. 
\label{fig:FoM} }
\end{wrapfigure}

With its measurements of the CMB {\it Planck\/} gave percent level constraints on seven $\Lambda$CDM parameters. Significantly more information can be extracted by a more sensitive survey that has better angular resolution.
Figure~\ref{fig:FoM} shows the increase in the figure of merit (FOM) since {\it COBE\/} for the $\Lambda$CDM model (dark purple) and several extensions. Our proposed survey, like that of the \PRISM\ mission concept \citep{PRISM2014I} from which it is inspired, could almost reach the cosmic-variance limit. It would outperform
what is expected from the combination of upcoming instruments, such as {\it LiteBIRD\/} for large angular scales \cite{Hazumi:2019lys},
and Simons Observatory \cite{Ade:2018sbj} or CMB-S4 \cite{Abazajian:2019eic} at higher angular resolution. 

\newpage

\subsection{Gravitational waves and inflation}

\vspace{-0mm}

CMB $B$-mode polarization is a unique window for detecting gravitational waves from inflation at an energy scale approximately a trillion times higher than those probed by the Large Hadron Collider. Similar to {\it PICO\/} \citep{2019arXiv190210541HB}, the proposed survey could also reach $\sigma (r) \sim {\cal O}(10^{-4})$, approximately 300 times better than the current sensitivity \cite{bk15,Akrami:2018odb} and an order of magnitude below the uncertainty targeted by any single experiment in the next decade. Classes of inflationary models motivated by string theory or supergravity, predicting $r \lesssim 10^{-3}$, could be probed unambiguously. As an example,  the Kahler geometry of $\alpha$-attractor models 
motivated by maximal supersymmetry~\cite{2016PhRvD..94l6015F,2017JHEP...04..144K}, could be probed entirely at high statistical significance.
In the case of no detection, a vast class of large-field inflationary models will be ruled out.

The survey will measure the power spectrum of curvature perturbations with a combination of range of angular scales and precision that are unprecedented for a single experiment. The forecast uncertainty on the scalar spectral index $\sigma(n_{\mathrm s}) \lesssim 0.0015$ is more than a factor of three tighter than current measurements, and will reach the required precision to constrain the reheating stage after inflation for a given inflationary model. The forecast precision on the running of the spectral index is 0.0015, at the same level as the theoretical predictions for single-field slow-roll inflationary models that provide a best-fit to \Planck data, such as $R^2$ or Higgs inflation. The proposed survey will therefore have the capability to discriminate among different inflationary models also on the basis of the shape of the curvature of the power spectrum.

Standard single-field slow-roll inflationary models predict primordial fluctuations with highly Gaussian statistics, compatible with the most recent \Planck\ constraints on the local, equilateral, and orthogonal shapes of the bispectrum \cite{Akrami:2019izv}. We can improve by a factor 2--3 on these bispectrum constraints, which are important to constrain models beyond the simplest ones, such as those with a non-trivial sound speed for the inflaton or with multiple fields. An enhanced sensitivity to the local shape of the bispectrum, down to $\sigma (f_{\mathrm{NL}}^\mathrm{local}) \sim 1$, an important threshold for multi-field inflationary models, can be reached by a tomographic cross-correlation of the lensing potential with deep radio or photometric surveys in preparation, such as EMU, SKA or LSST.

\subsection{Neutrinos and extra relics}

Inferring the neutrino mass sum $M_\nu$ from cosmological data will remain a crucial target in the long term, since planned laboratory experiments are not sensitive to the minimal value $M_\nu = 0.06$\,eV. Besides, it is important to exploit the synergy between cosmological surveys and laboratory searches, which are sensitive to different neutrino-related parameters and assumptions. On the cosmology side, precise measurements of $M_\nu$ require an exquisite mapping of both CMB anisotropies and large-scale structures (LSS). The two categories of observables are directly sensitive to the reduction in the growth rate of matter fluctuations induced by $M_\nu$, which CMB surveys probe through CMB lensing. CMB surveys will also play an essential role in accurately measuring other parameters like $\tau$, $n_{\rm s}$, $H_0$, and $\omega_\mathrm{c}$, that reduce degeneracies with $M_\nu$ in the analysis of LSS data.  Our new survey alone will reach a sensitivity of $\sigma(M_\nu)\simeq 0.04$~eV, and will be crucial in order to obtain $\sigma(M_\nu)\simeq {\cal O}(10^{-2})$~eV in combination with future galaxy, cosmic shear, and intensity-mapping surveys.

A plethora of extensions of the standard model of particle physics 
predict a relic density of extra light particles that would show up as an increase in the effective neutrino number $N_\mathrm{eff}$ beyond its standard value of 3.046 \cite{2016JCAP...07..051D}. Measuring $N_\mathrm{eff}$ is thus crucial for particle physics.
CMB anisotropies are the most sensitive probe of $N_\mathrm{eff}$.
The proposed survey will provide unprecedented sensitivity to $N_\mathrm{eff}$, with $\sigma(N_\mathrm{eff})=0.022$ using temperature and polarization, and $\sigma(N_\mathrm{eff})=0.016$ in combination with lensing extraction. In absence of extra relics, the standard value 3.046 will be distinguished from 3.0 at the 2--$3\sigma$ level, which will offer an accurate test of the standard model of neutrino decoupling and electron-positron annihilation. The possibility that any new scalar boson decouples from the standard model at some temperature $T<10^3$\,TeV will be either established or excluded at the 1.5$\sigma$ level (2$\sigma$ or 3$\sigma$ for a fermion or vector boson, respectively). A measurement compatible with 3.046 would prove with the same significance that no new particles have left thermal equilibrium between the decoupling of top quarks (at redshift $z\sim 10^{14}$) and today.  

Our survey will also be very sensitive to additional effects caused by the small mass of possible light non-thermal sterile neutrinos (whose effect would be roughly equivalent to a combination of $M_\nu$ and $N_\mathrm{eff}$), or to non-standard interactions in the neutrino sector (that would modify the so-called neutrino drag effects, particularly visible on intermediate and small scales in the polarization spectrum).

\subsection{Requirements and goals for primary CMB science}

Several projects plan next-generation observations of CMB anisotropies, in particular primordial polarization B-modes. Among those, the {\it LiteBIRD\/} satellite \cite{Hazumi:2019lys} has been selected as JAXA's strategic L-class mission for a launch in 2027, and the CMB-S4 experiment \cite{Abazajian:2019eic} was approved for CD-0 by the DOE in the US and is awaiting NSF participation.

We foresee that the {\it LiteBIRD\/} and CMB-S4 surveys can be improved with a subsequent space mission in the following ways:
\begin{enumerate}[nolistsep]
    \item full-sky maps with sensitivity and angular resolution matching the CMB-S4 3\% sky patch;
    \item isotropic maps (with no filtering along the scans);
    \item capability of full-sky delensing;
    \item capability to measure B-mode polarization with a sensitivity to $r$ of the level ${\cal O} (10^{-4})$;
    \item capability to de-lens with different methods (from CMB and from CIB maps);
    \item extended frequency coverage, in the sub-mm domain and between atmospheric windows.
\end{enumerate}

\noindent\myuline{\it Sensitivity and angular resolution:}\quad The combination of {\it LiteBIRD\/} and CMB-S4 will reach an aggregated CMB sensitivity of $\simeq 2.5 \, \mu$K.arcmin at $\simeq 1^\prime$ angular scale over 70\% sky, and a $\simeq 1\, \mu$K.arcmin sensitivity at $\simeq 1^\prime$ angular resolution in the 3\% sky deep patch (although with uneven frequency coverage, and possibly anisotropic filtering of the maps from the ground-based instruments).
With $\sigma_r \simeq 0.001$, both CMB-S4 and {\it LiteBIRD\/} would detect tensor modes at more than $5\sigma$ if $r>0.005$. 
We propose a full-sky survey with an order of magnitude improvement in sensitivity ($\sim$ 10 times lower noise spectrum, and $\sigma_r \sim 0.0001$), for redundant capability to detect $r \sim 0.001$ at more than 5$\sigma$ by internal delensing. An angular resolution of $\sim 5^\prime$ is adequate for most of the CMB science. However, there is added value to
increased angular resolution $\sim 1^\prime$ to deliver an unprecedented measurement of the CMB damping tail in temperature and polarization, which would lead to a sensitivity to neutrino physics beyond the experiments of the next decade.   

\noindent\myuline{\it Frequency range and number of channels:}\quad The survey should allow for near-full-sky foreground cleaning in both temperature and polarization. As argued above, and demonstrated with simulations in the \CORE\ study \citep{2018JCAP...04..023R}, a sensitive polarized imager with $\sim$\,20 frequency channels spanning a decade in frequency (from 60 to 600\,GHz) or more is adequate for this task. However, simulations have shown that a wider frequency range improves the effectiveness of some component separation techniques. We hence follow the \PICO\ design and target a frequency range from 20 to 800 GHz, as for the \PICO\ study \citep{2019arXiv190210541HB}.

\noindent\myuline{\it Control of Systematics and redundancy:}\quad High-sensitivity observations of CMB polarization require exquisite control of systematic effects. Space offers the best environment for this. Methods for systematics control have been developed and assessed in the context of the \CORE\ study \citep{2018JCAP...04..014D}.

\noindent
\begin{center}
\fbox{\begin{minipage}[h][][t]{0.98\textwidth} 
\textbf{Requirements and goals for primary CMB science} \\
Full-sky CMB sensitivity of $\simeq 1\,\mu $K.arcmin;\\
Angular resolution $5^\prime$ requirement, and goal $1^\prime$;\\
$\simeq 20$ frequency channels in the 20--800\,GHz frequency range;\\
Demonstrated control of systematics effects.
\end{minipage}}
\end{center}

\vspace{-0.1in}

\subsection{Tests of homogeneity and isotropy}

Testing the apparent large-scale anomalies observed in CMB temperature maps calls for investigations of our Universe on the largest scales using other observables than $T$ \cite{2019arXiv190602552P} -- our mission can do this with many different probes of the Hubble volume. An intriguing possibility is to measure the dipole of cosmic backgrounds other than the CMB to test that it is entirely due to our motion, as usually assumed \citep{2018JCAP...04..021B}.
Other possible tests of the statistics on scales comparable to the Hubble radius include the stationarity and large-scale modulation of lensing potential maps, CIB fluctuations, SZ cluster counts, and the measurement of local quadrupoles at various redshifts through the polarised SZ effect. A survey capable of measuring these signals would be unique in assessing the large-scale homogeneity and isotropy of our Universe.  These go beyond tests based on CMB temperature and polarization maps, which probe our Universe in a single redshift shell, and are limited by cosmic variance.


\section{Information from the CMB's near-blackbody spectrum}
\label{sec:distortions}

The precise shape of the CMB energy spectrum encodes new information that can be extracted using absolute CMB spectroscopy.
At redshifts $z\gtrsim 2\times 10^6$, thermalization processes are efficient and promptly restore a near perfect blackbody spectrum of the CMB if full thermal equilibrium was perturbed. However, at later epochs, starting few months after the Big Bang, 
traces of energy-releasing or photon-injecting processes can be found by measurements of departures from a perfect blackbody spectrum.
While classically CMB spectral distortions are described as a sum of $\mu$- and $y$-type distortion signals~\citep{Zeldovich1969, Sunyaev1970mu, Burigana1991, Hu1993}, modern treatments of the problem have demonstrated that far more than just two numbers can be extracted~\citep[e.g,][]{Chluba2011therm, Khatri2012mix, Chluba2013fore, Chluba2013PCA}.
Measurements can constrain processes expected within $\Lambda$CDM including the damping of primordial perturbations and the recombination radiation, and open discovery space to the pre-recombination Universe which cannot be accessed directly any other way.
{\it COBE}-FIRAS still defines the long-standing benchmark for CMB spectral distortions, but several orders of magnitude of sensitivity improvements are in principle possible, as envisioned for \PIXIE~\citep[e.g.,][]{Kogut2011PIXIE}, the spectrometer of \PRISM~\citep{PRISM2014II} and \SPIXIE~\citep{Kogut2019WP}. 

\subsection{Spectral distortions as a new test of inflation}

Fluctuations set up by inflation dissipate their energy through photon diffusion. This causes a CMB distortion signal that can be used to derive stringent constraints on the amplitude and shape of the primordial power spectrum at scales inaccessible to other probes \citep{Sunyaev1970diss, daly1991, Chluba2012, Chluba2012Inflaton}.
A spectrometer like \PIXIE could rule out excess power at wavenumbers $k\simeq 50\,{\rm Mpc}^{-1}$--$10^4\,{\rm Mpc}^{-1}$ at the level of $P(k)\gtrsim 10^{-8}$ \citep{Chluba2012Inflaton}. This would place novel constraints on a wide range of early-Universe models outside of standard slow-roll inflation, including features or inflection points in the potential, particle production and waterfall transitions \citep{Chluba2019WP}. 
With 10 times better sensitivity (a fraction of a Jy/sr) a spectrometer could detect the expected $\Lambda$CDM $\mu\simeq 2\times 10^{-8}$ at the $3\sigma$-level~\citep{Chluba2019WP}; see Figure~\ref{fig:SD-and-FG}. This would constrain inflation models, with a guaranteed target within standard slow-roll inflation. In combination with a future CMB imager, a spectrometer could improve the limits on the running of the spectral index by a factor of about 2  \citep{PRISM2014II, Chluba2019WP}. 
Primordial local-type non-Gaussianity could also be constrained using $\mu$-distortion anisotropies \citep{Pajer2012, Ganc2012, Remazeilles2018mu}, providing an independent new probe of early-Universe physics \citep{Chluba2019WP}.

\subsection{Reionization and structure formation}

\begin{figure}
\vspace{-1.5cm}
\hspace{-6mm}
\parbox{10cm}{\centerline{
\includegraphics[width=10
cm]{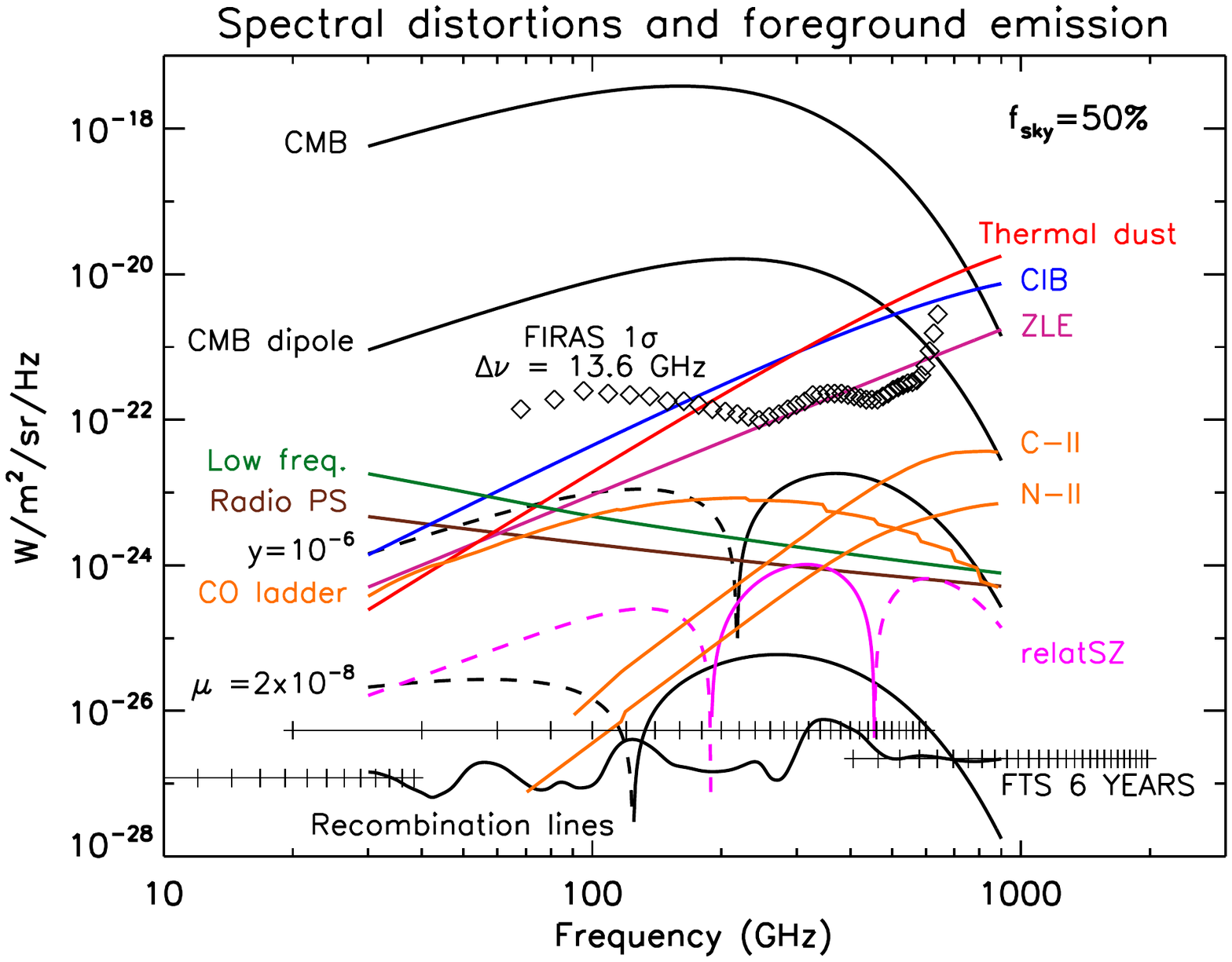} } }
\vspace{-6mm}
\parbox{7.0cm}{
\caption{\captiontext The signals of interest that contribute to the total difference of microwave emission with a perfect blackbody span about 8 orders of magnitude in amplitude. The survey proposed here (indicated by black horizontal lines with vertical bars at the central frequencies) improves upon COBE-FIRAS (diamonds) by about four orders of magnitude, and would access the overall $y$-parameter signal (black dash and solid) discussed in section 2.1 and 5.2, as well as some of the signal from atomic and molecular lines discussed in section 3.3. It may make a few $\sigma$ detection of the $\mu$-distortion (black dash and solid) expected in $\Lambda$CDM and of the rSZ effect (magenta dash and solid), assuming that the total foreground emission can be measured and subtracted at the same level of accuracy. 
\label{fig:SD-and-FG}}}
\vspace{-1.3in}
\end{figure}

The largest $\Lambda$CDM distortion is created by the low-$z$ structure formation and reionization process \citep{Cen1999, Refregier2000, Miniati2000, Oh2003}. The first stars, accreting black holes, and shocks heat the baryons and electrons, which then up-scatter CMB photons to create an average $y$-type distortion. The overall expected distortion is $y\simeq$~few~$\times10^{-6}$ \citep{Refregier2000, Hill2015}, one order of magnitude below the upper bound from {\it COBE}-FIRAS. As shown in Fig.~\ref{fig:SD-and-FG}, this could be constrained to the sub-percent level with a future mission \citep[see also][]{Chluba2019WP}. 

A large part of the low-redshift Compton-$y$ signal is due to halos with masses $M\simeq 10^{13}\,{\rm M}_\odot$, which contain virialized gas with an electron temperature of $k T_{\rm e}\simeq 2$--$3\,$keV. This causes a relativistic temperature correction (rSZ) \citep{Wright1979, Sazonov1998, Itoh98} that can directly tell us about feedback mechanisms \citep{Hill2015}. Both the $y$ and the rSZ distortion depend directly on the shape and amplitude of the halo mass function, providing another cosmological measure of the growth of structure. With sufficient sensitivity, the survey could determine the average relativistic temperature with S/N of tens -- assuming foregrounds can be controlled -- and constrain feedback physics that currently are still very poorly understood \citep{Chluba2019WP}. 
A direct measurement of the average rSZ temperature would also shed new light on the ``missing baryon problem'' \citep{Cen1999} without the need to resolve the warm-hot-intergalactic medium.
Measurements at $\nu\gtrsim 500$\,GHz will probe the total cosmic-ray energy density of the Universe through the non-thermal relativistic SZ effect \citep{Chluba2019WP}.
Furthermore, extremely precise spectrum measurements down to $\sim 10$\,GHz will allow us to greatly improve our knowledge of the diffuse free-free emission associated with cosmological reionization, distinguishing between various models \citep{1999ApJ...527...16O,2011MNRAS.410.2353P, 2014MNRAS.437.2507T}, and will shed light on the controversial question of a potential low-frequency background temperature excess raised by ARCADE 2 and EDGES results \citep{2011ApJ...734....6S,EDGESobs2018Nature}.  
These illustrate some of the unique opportunities in CMB spectroscopy.

\subsection{Probing dark matter and particle physics}

Dark matter is another example of how spectral distortions allow us to probe new physics. Non-baryonic matter constitutes $\simeq25$\% of the energy density of the Universe, but its nature remains unknown. The long-favored WIMP-scenario is under increasing pressure \citep{Ahmed:2009zw, Aprile:2012nq, Angloher:2015ewa, Agnese:2015nto, Tan:2016zwf, Akerib:2016vxi}, and emphasis is gradually shifting towards alternatives, prominent examples being axions, sterile neutrinos, sub-GeV DM or primordial black holes \citep{Jungman1996, Feng2003PhRvL, Feng2003, Kusenko2009, Feng2010, Carr2010, Marsh2016Rev}.
To solve this puzzle, a coordinated multi-tracer approach that combines different particle physics and cosmological probes is needed. 

Measurements of the CMB anisotropies themselves have clearly helped to establish the presence of DM on cosmological scales and provided tight constraints on DM annihilation and decay \citep{Ellis1992, Adams1998nr, Chen2004, Padmanabhan2005, Galli2009, Slatyer2009, Slatyer2017, Poulin2017} and DM-SM-interactions \citep{Wilkinson2014, Dvorkin2014, Wilkinson2014b, Gluscevic2018}. However, for DM annihilation and decay CMB anisotropies quickly lose constraining power before recombination ($z\gtrsim 10^3$), being impeded by cosmic variance. 
Similarly, measurements of light-element abundances \citep{Ellis1992, Kawasaki2005, Jedamzik2008, Kawasaki2018}, which are only sensitive to non-thermal energy release above nuclear-dissociation thresholds in the pre-recombination era \citep{Chluba2013PCA, Poulin2015Loop}, saturated their limits due to astrophysical uncertainties. {\it This is where CMB spectral distortions offer a valuable complementary probe.} 
For decaying particle scenarios, distortions are sensitive to particles with lifetimes $t\simeq 10^6$--$10^{12}\,{\rm s}$ 
\citep{Sarkar1984, Ellis1985, Kawasaki1986, Hu1993b, Chluba2011therm, Chluba2013PCA, Aalberts2018}, providing a direct measurement of particle lifetimes via residual distortions \citep{Chluba2013fore, Chluba2013PCA}. Similarly, annihilating particles can be constrained using distortions: $\mu$-distortions are sensitive to light particles ($m \lesssim 100$ keV) and complement $\gamma$-ray searches for heavier particles \citep{McDonald2001, Chluba2013fore}. The rich spectral information added by various non-thermal processes \citep{Liubarskii83, Chluba2008c, Chluba2010a, Chluba2015GreensII, Slatyer2015, Acharya2018} will allow us to glean even more information about the nature of dark matter. 

This is new territory and
more work is required; however, it is already clear that CMB spectral distortions can meaningfully probe scenarios involving axions \citep{Tashiro2013, Ejlli2013, Mukherjee2018}, gravitino decays \citep{Ellis1985, Dimastrogiovanni2015}, strings \citep{Ostriker1987, Tashiro2012b}, DM-SM-interactions \citep{Yacine2015DM, Diacoumis2017, Slatyer2017}, macroscopic DM \citep{Kumar2018}, and primordial magnetic fields \citep{Jedamzik2000, Sethi2005, Kunze2014, Wagstaff2015}. 
A CMB spectrometer that reaches the level of $\mu\simeq 10^{-8}$ after foreground marginalization can rule out a vast class of particle-physics models and also allow a first detection of the guaranteed $\mu$-distortion signal from the damping of primordial acoustic modes.

\subsection{The cosmological recombination radiation}

The cosmological recombination process causes another small but inevitable distortion of the CMB. Line emission from hydrogen and helium injects photons into the CMB, which after redshifting from $z\simeq 10^3$ are visible today as complex frequency structure in the microwave bands \citep{Dubrovich1975, RybickiDell94, DubroVlad95, Kholu2005, Wong2006, Jose2006, Chluba2006, Jose2008, Yacine2013RecSpec}. The cosmological recombination radiation (CRR) has a simple dependence on cosmological parameters and the dynamics of recombination; since it includes not only hydrogen but also helium recombinations, it probes eras beyond the last-scattering surface observed by CMB anisotropies \citep{Chluba2008T0, Sunyaev2009, Chluba2016CosmoSpec}. The signals are however weak and require noise levels of the order of 0.1 Jy/sr or better. 

\subsection{Requirements and goals for spectral distortion science}

\noindent\myuline{\it Sensitivity and angular resolution:}\quad CMB spectral distortion science is new territory, with a vast potential for discovery, but many unknowns on the path. 
For this reason, we set a relatively safe sensitivity requirement in terms of accessible science, i.e. an aggregated sensitivity at the level of $10^{-25}$\,W/m2/sr/Hz (10\,Jy/sr), sufficient to detect the mean $y$ level and high-redshift spectral lines plotted in Fig. \ref{fig:SD-and-FG}. As a goal, we target 100 times better sensitivity, to look also for  the faintest spectral distortion signals down to the CRR. An angular resolution of $\simeq 1^\circ$ is sufficient for selecting data from relatively clean regions of the sky in terms of galactic and zodiacal foreground contamination.

\noindent\myuline{\it Frequency range and spectral resolution:}
The survey must cover the region where the specific spectral signatures of the  distortions make them distinguishable from other emissions. We require 30-600\,GHz coverage, with a goal of 10-2000. A spectral resolution corresponding to $R\simeq 10$ is adequate to distinguish the various components.

\noindent
\begin{center}
\fbox{\begin{minipage}[h][][t]{0.98\textwidth} 
\textbf{Requirements and goals for CMB spectroscopy}

Sensitivity requirement: 10 Jy/sr aggregated, full mission; Goal: 0.1 Jy/sr;\\
Angular resolution $\simeq 1^\circ$;\\
Frequency coverage from 30 to 600 GHz (requirement), with a goal of 10-2000\,GHz;\\
Spectral resolution (frequency channel width) in the 2-60 GHz range.
\end{minipage}}
\end{center}

\section{Possible mission profiles}


Summarizing the requirements detailed above, this science program requires detecting the following signals with high signal-to-noise ratio and high precision, over the entire sky.
\begin{enumerate}[nolistsep]
    \item Thermal SZ emission from most galaxy clusters in the Hubble volume, to map hot ionized gas in the cosmic web: angular resolution 1.5 to $1^\prime$; CMB sensitivity $\Delta_y \simeq 10^{-6}$ at 1$\sigma$ per arcmin pixel around 150 and 350\,GHz; $\simeq 20$ frequency channels in the 50--800\,GHz frequency range.
    \item CMB anisotropies generated by lensing effects and the kSZ effect; angular resolution 1.5 to $1^\prime$; CMB sensitivity $\Delta_T \simeq 1$ to $0.6\,\mu$K.arcmin; $\sim 20$ frequency channels covering the 50--800\,GHz frequency range.
    \item CMB anisotropies from $z\simeq 1100$; angular resolution $5^\prime$ to $1^\prime$; sensitivity $\Delta_T \simeq 1\,\mu$K.arcmin; $\sim 20$ frequency band in the 20--800\,GHz frequency range.
    \item Absolute emission and fluctuations from dust continuum and [{\sc Cii}]/CO lines across a wide range of redshifts (up to $z\simeq 10$) with spectral resolution $R=300$ extending from $\sim 100$ to $\sim 1000$\,GHz; angular resolution 1--$5^\prime$; capability to map deep patches.
    \item Absolute spectrum of the microwave sky emission from 10 to 2000\,GHz; angular resolution $\sim 1^\circ$; sensitivity integrated over the full observing time in the 0.1--10\,${\rm Jy}\,{\rm sr}^{-1}$ range.
\end{enumerate}

Overall, the goal would be to achieve, with a combination of instruments, a spectro-polarimetric survey of the entire sky from 10 to 2000\,GHz, with angular resolution $1^\prime$ to $1.5^\prime$, and sensitivity matching the requirements of the above science goals.

\subsection{Mission overview}

The space mission should perform those necessary observations that cannot be done better from the ground. The key design elements are below. \\
$\bullet$ Angular resolution between $1.5^\prime$ and $1^\prime$ at $\geq300$\,GHz. This requires a telescope with aperture between 2.8\,m (requirement) and 4.2\,m (goal). We rely on ground-based telescopes for the smallest scales at lower frequencies. A larger telescope in space would be challenging and not cost-efficient. \\
$\bullet$ Two focal-plane instruments: a broad-band polarized imager from 20 to 800\,GHz for CMB anisotropies, tSZ, kSZ, lensing, and a spectrometer with $R\simeq 300$ for CIB tomography and line intensity mapping from 100 to 1000\,GHz.\\
$\bullet$ A set of FTSs covering 10--2000\,GHz for measurement of the absolute spectrum.

This requires an L-class mission, or a combination of L and M missions. Downscoping to M-class would require reducing the telescope size by a factor $\sim 3$, and possibly relaxing the temperature requirement for the primary mirror, for a space mission similar to \CORE\ or \PICO. This would significantly degrade most of the high resolution science (cluster and protocluster surveys, CMB lensing science, velocity flows from kSZ), but remains appealing for primary CMB science (primordial gravitational waves, cosmological parameters constraints), as shown in the context of the \CORE\ and \PICO\ studies. Such a mission could be envisaged in the early 2030s in the context of an international collaboration, e.g. between NASA and ESA.

\subsection{Instruments}
\label{sec:instruments}

\begin{figure}
\hspace{-0.1in}
\parbox{4.3in}{
\includegraphics[width=4.3in]{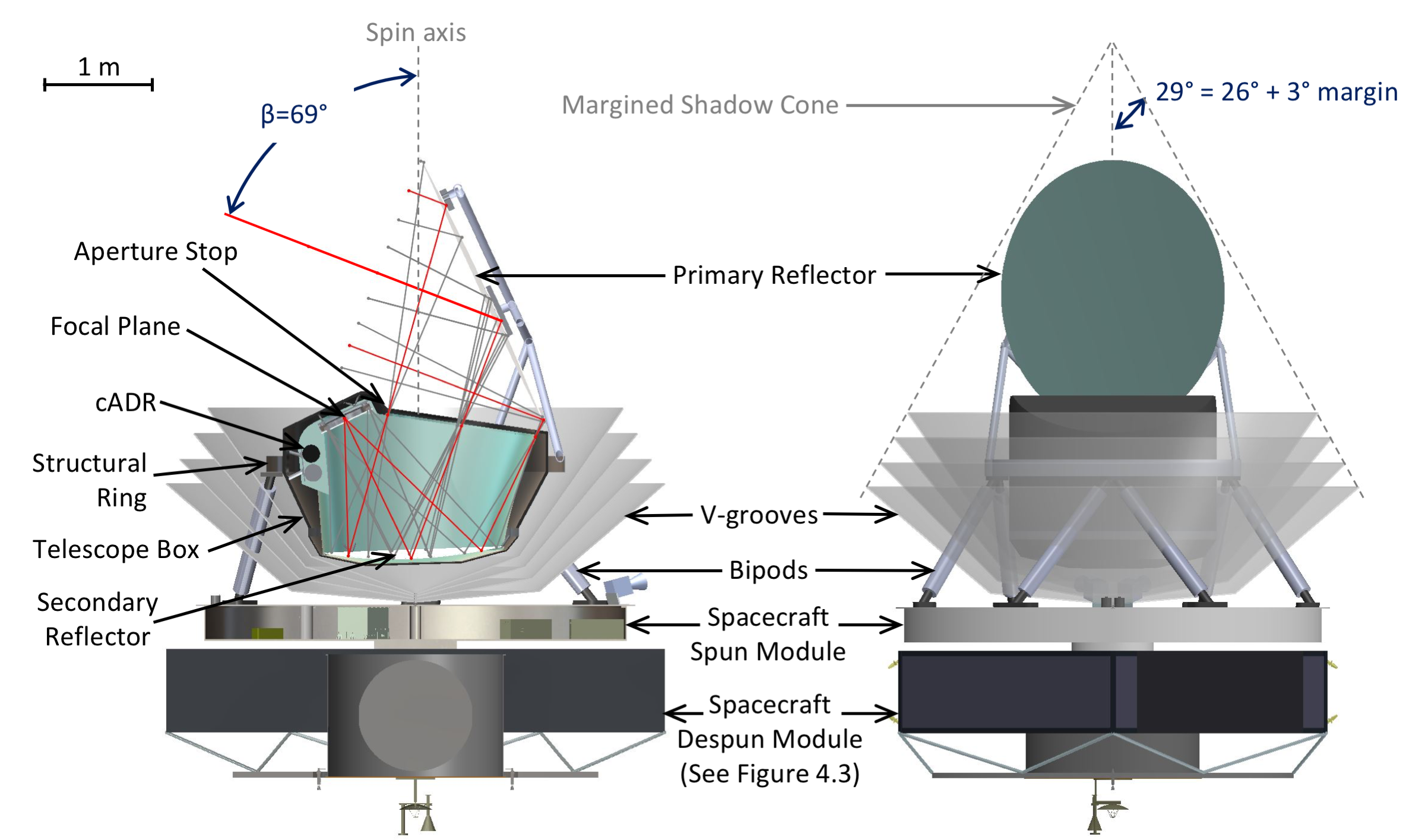} }
\parbox{2.2in}{
\includegraphics[width=2.2in]{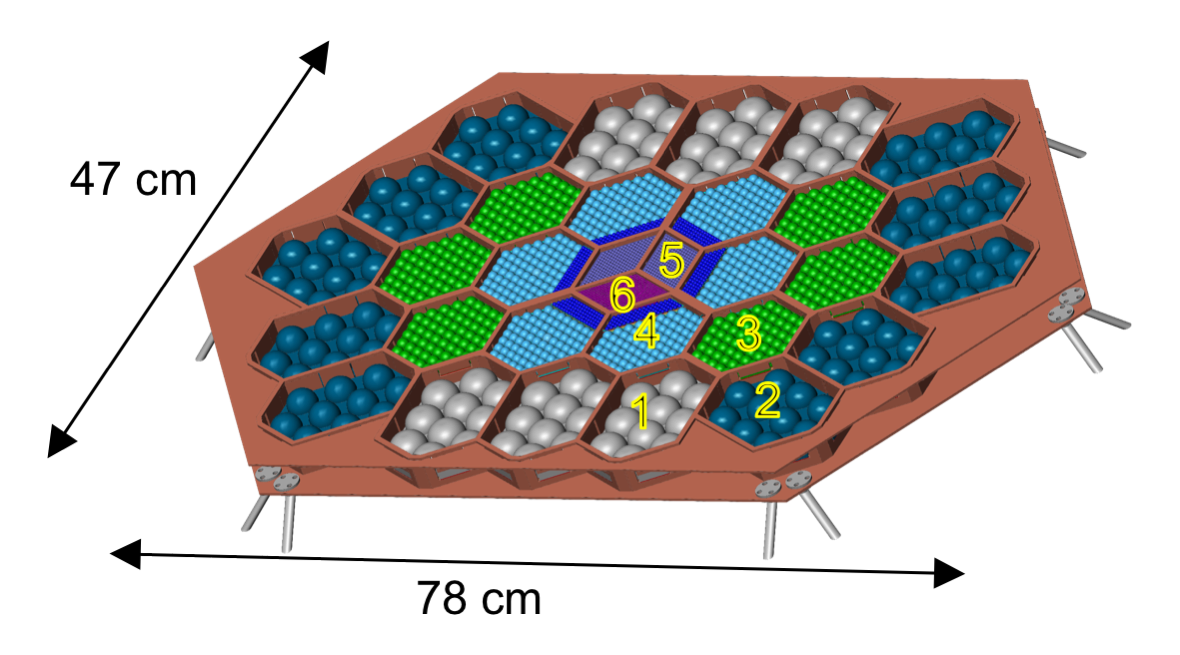} }
\vspace{-0.1in}
\caption{\captiontext
\PICO\ overall configuration in side view and cross section (left), front view with V-Groove assembly shown semi-transparent (middle), and the focal plane (right) (reproduced with permission from~\citet{2019arXiv190210541HB}). A 2.8-m entrance aperture is achievable by scaling the \PICO\ optical design by a factor of two and using deployable shields. The \PICO\ focal plane, which containts 12,996 TES bolometers, achieves the noise levels baselined for the survey proposed here. 
\label{fig:InstrumentCAD}} 
\vspace{-0.25in}
\end{figure}

\noindent\myuline{\it Polarimetric imager:}\quad The polarimetric imager must observe the polarized emission in several broad bands ($\Delta\nu/\nu \simeq 0.25$) covering the frequency range at which CMB anisotropies and SZ effects dominate. Its science goals can be achieved with an instrument that is a straight-forward extension of today's technologies, specifically based on the design for \PICO, an instrument proposed for consideration by the US-2020 decadal panel~\citep{2019arXiv190210541HB}; see Fig.~\ref{fig:InstrumentCAD}. Table~\ref{tab:polarised-imager} gives the expected performance for a full-sky survey and a deep patch, observed for 2 years and 6 months, respectively, using this instrument. 

The current telescope design for \PICO, which has a 1.4-m diameter aperture, allows for a factor 2 increase to a diameter of 2.8\,m and the instrument would fit with no other changes (except for the need of deployable shields) within the Ariane~6 shroud.  Changes to the optical design should allow increase to diameters between 3.5 and 4\,m. The {\it Herschel} mission had a 3.5-m telescope. 

The focal plane is continuously maintained at 0.1\,K. Several technologies including continuous adiabatic-demagnetization refrigerators~\citep{Shirron2012,Shirron2016} and continuous-cycle dilution refrigerators~\citep{brien2018} are either already near-mature, or should be mature by the 2030s.
The focal plane module contains four elements: (1) low-pass filters rejecting electromagnetic (EM) radiation above the highest band ($\sim850\,$GHz); (2) monolithic arrays of thousands of pixel elements that couple the EM radiation from space to transmission lines, which then channel the power to transition-edge-sensor (TES) bolometers converting the deposited power to current signals; and (3) front-end SQUID amplifiers. Current technologies allow coupling a broad-band of EM radiation into a focal plane pixel using broad-band antennas or horns~\citep{Suzuki2014, nist_design}, and then channeling specific frequency bands into their bolometers using on-wafer filters. We will use these technologies with up to three bands per pixel for frequencies up to $\sim450\,$GHz, which is close to the Nb bandgap. At higher frequencies, for which we cannot use superconducting Nb to channel the EM into transmission lines, we will use direct absorption onto polarization-sensitive bolometers with one frequency band per pixel. 

Both antenna-based and horns-based EM coupling of the radiation are polarization preserving; they do not alter the incident polarization, but can select for detection one of the polarization states. End-to-end polarimety is achieved by splitting the incident radiation into two orthogonal polarization states, and arranging the focal plane to have sensitivity to at least three orientations~\citep{2019arXiv190210541HB}.  

The readout is based on the multiplexing $N$ detectors onto two readout lines (where $N$ is called the multiplexing factor). Both time-domain-based and frequency-domain-based multiplexing are in use by operating experiments. For PICO we assumed a conservative factor of 128. Systems with multiplexing factors between 2000 and 4000 are in development. 

\vspace{-0.05in}

\begin{footnotesize}
\begin{center}
\begin{table}[]
    \caption{Performance forecast for the polarized imager, from estimated performance of the instrument in the {\it PICO\/} study \citep{2019arXiv190210541HB}, and 3.5-m aperture optics. The first set of sensitivity columns are for a 2-year full-sky survey, the next five for a deeper patch of 5\% sky observed for a total of 6 months. Sensitivities for $y$ in the negative part of the tSZ spectrum are conventionally noted with a negative sign. CMB, $y$, and brightness sensitivities are for $1^\prime$ pixels, while flux sensitivities (in ${\rm W}\,{\rm m}^{-2}$) and point source sensitivities (in mJy) are integrated in the beam. Polarization sensitivities are obtained by multiplying these numbers by $\sqrt{2}$. The last line gives the aggregated focal plane array sensitivity to signals with the color of CMB or tSZ (actual sensitivity will be reduced after separation of the astrophysical components).}
    \vspace{2mm}
    \footnotesize   
    \centering
    \scalebox{1}{%
    \begin{tabular}{|c|c|c c c c|c c c c c|}
    \hline
    $\nu$ & Beam & CMB $\sigma_I$  & tSZ $\sigma_y$ & $\sigma_I$ & PS 5$\sigma$ & CMB $\sigma_I$  & tSZ $\sigma_y$ & $\sigma_I$ & Flux $\times 10^{-20}$ & PS 5$\sigma$ \\
    (GHz) & (arcmin) & ($\mu$K) & ($\times 10^6$) & (${\rm kJy}\,{\rm sr}^{-1}$) & (mJy) & ($\mu$K) & ($\times 10^6$) & (${\rm kJy}\,{\rm sr}^{-1}$) & (${\rm W}\,{\rm m}^{-2}$) & (mJy) \\
    \hline
    \hline
21 & 15.36 & 18.41 & $-3.41$ & 0.24 & 27.89 & 8.23 & $-1.52$ & 0.11 & 13.1 & 12.47 \\
25 & 12.8 & 12.88 & $-2.4$ & 0.24 & 19.12 & 5.76 & $-1.07$ & 0.10 & 10.69 & 8.55 \\
30 & 11.32 & 8.74 & $-1.64$ & 0.23 & 14.51 & 3.91 & $-0.73$ & 0.10 & 9.73 & 6.49 \\
36 & 9.44 & 6.13 & $-1.16$ & 0.23 & 10.09 & 2.74 & $-0.52$ & 0.10 & 8.12 & 4.51 \\
43 & 8.88 & 6.13 & $-1.18$ & 0.33 & 12.56 & 2.74 & $-0.52$ & 0.14 & 12.08 & 5.62 \\
52 & 7.35 & 4.29 & $-0.84$ & 0.33 & 8.64 & 1.92 & $-0.37$ & 0.14 & 10.05 & 3.86 \\
62 & 5.12 & 4.14 & $-0.84$ & 0.44 & 5.57 & 1.85 & $-0.37$ & 0.19 & 7.72 & 2.49 \\
75 & 4.27 & 3.22 & $-0.68$ & 0.48 & 4.23 & 1.44 & $-0.30$ & 0.21 & 7.10 & 1.89 \\
90 & 3.8 & 2.14 & $-0.49$ & 0.43 & 3.01 & 0.96 & $-0.22$ & 0.19 & 6.06 & 1.34 \\
108 & 3.16 & 1.68 & $-0.43$ & 0.45 & 2.16 & 0.75 & $-0.19$ & 0.20 & 5.21 & 0.96 \\
129 & 2.96 & 1.68 & $-0.51$ & 0.57 & 2.39 & 0.75 & $-0.22$ & 0.25 & 6.91 & 1.07 \\
155 & 2.48 & 1.38 & $-0.56$ & 0.56 & 1.67 & 0.61 & $-0.25$ & 0.25 & 5.79 & 0.74 \\
186 & 1.72 & 3.06 & $-2.40$ & 1.42 & 2.02 & 1.37 & $-1.07$ & 0.63 & 8.42 & 0.90 \\
223 & 1.44 & 3.52 & 15.29 & 1.70 & 1.69 & 1.57 & 6.84 & 0.76 & 8.45 & 0.75 \\
268 & 1.28 & 2.3 & 1.05 & 1.02 & 0.8 & 1.02 & 0.46 & 0.45 & 4.83 & 0.36 \\
321 & 1.04 & 3.22 & 0.69 & 1.15 & 0.59 & 1.44 & 0.31 & 0.51 & 4.28 & 0.26 \\
385 & 1.00 & 3.52 & 0.46 & 0.84 & 0.4 & 1.57 & 0.20 & 0.37 & 3.47 & 0.18 \\
462 & 0.84 & 6.90 & 0.61 & 0.87 & 0.29 & 3.08 & 0.27 & 0.39 & 3.07 & 0.13 \\
555 & 0.60 & 35.29 & 2.24 & 1.81 & 0.31 & 15.78 & 1.00 & 0.81 & 3.89 & 0.14 \\
666 & 0.52 & 136.5 & 6.48 & 2.06 & 0.26 & 61.07 & 2.89 & 0.92 & 3.98 & 0.11 \\
799 & 0.44 & 807.1 & 29.41 & 2.43 & 0.22 & 360.96 & 13.15 & 1.08 & 4.03 & 0.10 \\
    \hline
    \hline
Total &  &       0.66 &       0.17 &  &  &       0.29 &      0.077 &  &  &  \\
    \hline
    \end{tabular}}
    \label{tab:polarised-imager}
\end{table}
\end{center}
\end{footnotesize}

\vspace{-0.45in}

\noindent\myuline{\it Absolute spectrophotometry:}\quad Order-of-magnitude improvements to current upper limits for CMB spectral distortions require continuous spectra at modest spectral resolution, covering 6 or more octaves in frequency with part-per-million channel-to-channel calibration stability. Fourier transform spectroscopy is ideally suited to this task.
The FTS measures the difference spectrum between the sky and a blackbody calibrator. Unlike an imager, whose frequency channels are set by bandpass filters, the central frequency,
channel bandwidth, and channel-to-channel covariance of an FTS's synthesized frequency channels depend only on the sampling (apodization) of the interferograms and can be determined a priori. The photon noise to first order is the same for all channels; it depends on the integrated optical load over the total passband and scales linearly with the synthesized channel width.

A broad-band FTS based on the \PIXIE design \cite{Kogut2011PIXIE} would achieve the science goals outlined above. \PIXIE uses a single FTS with 15-GHz channels extending from 15\,GHz to 6\,THz. If foregrounds were negligible, \PIXIE could detect the $\mu$-distortion from Silk damping of primordial density perturbations at 2--3~$\sigma$ and detect recombination lines at comparable levels. Astrophysical foregrounds degrade the ideal performance and require additional sensitivity  at both low and high frequencies \citep{abitbol_pixie}.

Such sensitivity could be obtained using several nearly-identical FTS modules, each with different optical passbands and synthesized frequency channels, optimized for either the CMB distortion signals or for the measuring the competing foreground emission. Table~\ref{tab:fts} shows the performance for a design with three modules covering, respectively, low, middle, and high frequencies (inspired from \citep{Kogut2019WP}).
Each module is based entirely on existing technologies. 

\begin{footnotesize}
\begin{center}
\begin{table}[]
    \caption{Multi-module absolute spectrometer; The mission sensitivity in the last column assumes 70\% useful data and a 6-year mission.}
\footnotesize
    \centering
    \begin{tabular}{|l|c c c c |c|}
\hline
Module \hspace{2mm} & \hspace{2mm} $\nu_{\rm min}$ (GHz) \hspace{2mm} & \hspace{2mm} $\nu_{\rm max}$ (GHz) \hspace{2mm} & \hspace{2mm} $\Delta \nu$ (GHz)
\hspace{2mm} & \hspace{3mm} Sensitivity (Jy.$\sqrt{\rm s}$) \hspace{2mm} & \hspace{2mm} Mission sens. (${\rm Jy}\,{\rm sr}^{-1}$) \hspace{2mm}  \\
\hline
\hline
LFM & 9.6 & 38.4 & 2.4 & 1435 & 0.12\\
MFM & 20 & 600 & 20 & 6200 & 0.54\\
HFM & 406 & 2000 & 58 & 2520 & 0.22\\
  \hline
   \end{tabular}
    \label{tab:fts}
\end{table}
\end{center}
\vspace{-0.25in}
\end{footnotesize}

\vspace{-0.2in}

\begin{figure*}[htb]
\centering
\includegraphics[width=0.9\textwidth]{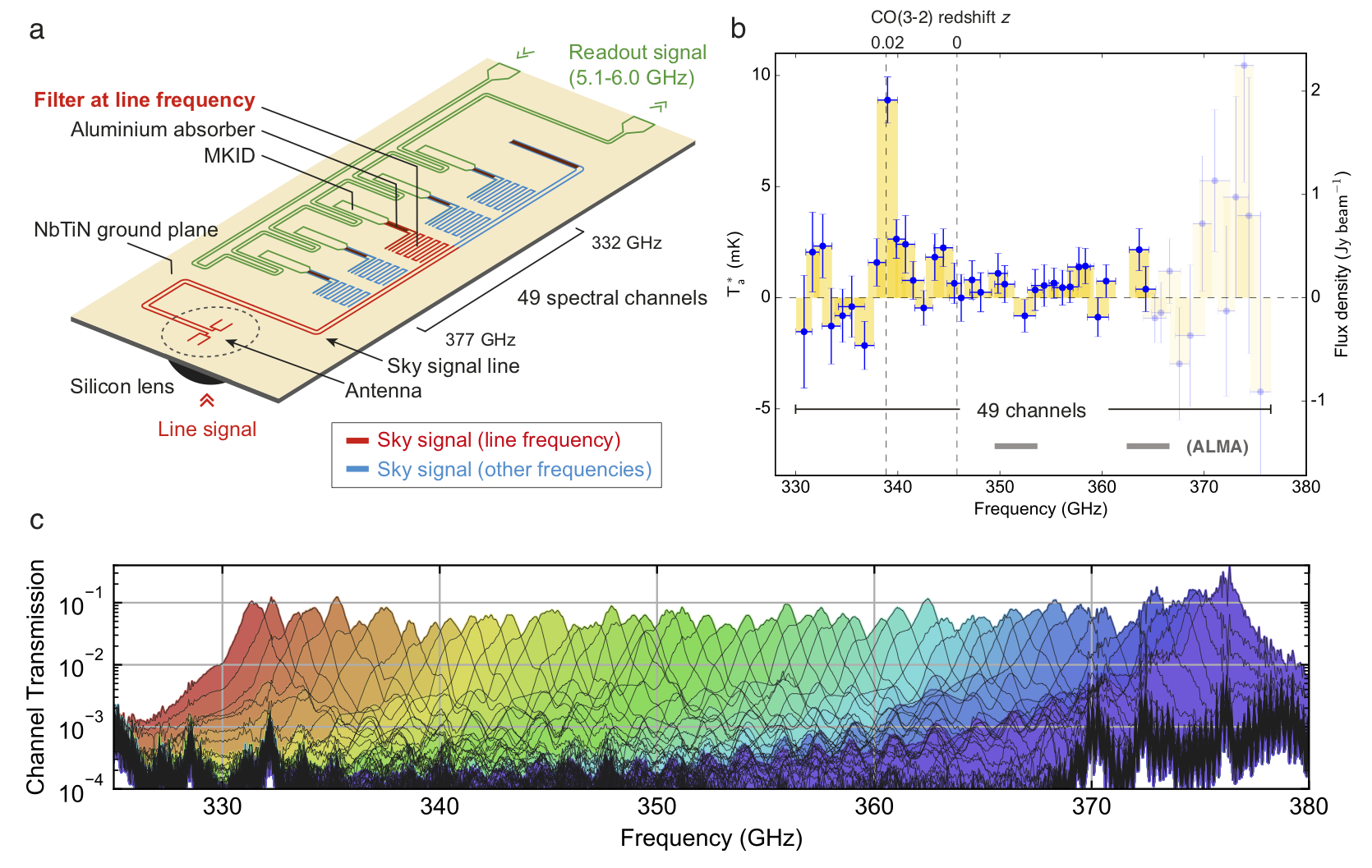}
\vspace{-0.25in}
\caption{\captiontext {\it (a)\/}: Sketch of the spectrometer chip. {\it (b)\/}: Measured spectrum from VV114 with 2017 prototype. {\it (c)\/}: Lab calibration of the spectral response of the individual filters. Reproduced with permission.~\citep{Endo2019}}  
\label{fig:Deshima}
\vspace{-0.15in}
\end{figure*}

\noindent\myuline{\it Filter-bank spectrometer:}\quad  The recently demonstrated on-chip filter-bank spectrometer is an ideal candidate to provide spectral filtering and radiation detection over large bandwidths with minimal weight and complexity \cite{Endo2019}. As shown in Fig.~\ref{fig:Deshima}, it consists of a chip fabricated from an NbTiN superconducting film that creates an electrical circuit which combines radiation detection by means of an antenna, spectral filtering by a filter-bank spectrometer and detection by using background limited MKID detectors and their readout \cite{Baselmans2017}. The technology proposed in Ref.~\cite{Endo2019} is intrinsically limited to a 90\,GHz to 1.1\,THz band due to the properties of the materials used, but can be easily upgraded to ultra-large bandwidths using leaky-wave antennas \cite{Bueno2017}. Developments using low-Tc superconductors (such as Ti or TiN) are needed to go down to 50\,GHz; a dielectric-based filter-bank, taking advantage of the low loss tangents of crystalline Si, is needed to extend the frequency range to 2\,THz.   

\noindent\myuline{\it Other instrumental options:}\quad Other options can be considered for achieving the proposed spectroscopic survey. An FTS in the focal plane could be implemented by means of a steerable mirror reorienting the beam towards the different instruments. This solution is attractive for  spectroscopy at high angular resolution, with the price of complexity for the focal plane assembly.
This solution is not well suited for very accurate measurement of CMB spectral distortions, which do not require high angular resolution, but must compare the sky to a movable (and tunable) calibrator, and require excellent stray-light control. In the focal plane of a large telescope, a planar FTS would be an interesting option, and easier to integrate close to other instruments. This kind of device, not ready yet, make use of superconducting-microstrip or coplanar delay lines, currently investigated by several groups. Being coupled to planar antennas, they would be polarization sensitive, and inherently single mode. For line spectroscopy a Fabry-Perot interferometer and a grating spectrometer would be interesting options. A cold grating coupled to an actively cooled telescope meets the condition of the lowest photonic background for sub-mm line search, but it size scales linearly with the spectral resolution.

\section{Scientific and technological roadmap}

\vspace{-0mm}
\subsection{Scientific heritage and complementarity of probes}
\vspace{-0mm}

The roadmap to the proposed mission benefits from intermediate projects that address a fraction of the science case.

$\bullet$ {\it LiteBIRD\/} plans to observe CMB polarization on large angular scales to search for inflation-produced gravitational waves. Our proposed survey is designed for other scientific objectives, but still improves upon the LiteBIRD target search limits by a factor of 10.

$\bullet$ The Simons observatory, and later CMB-S4, will observe the CMB and galaxy clusters with 1--1.5$^\prime$ angular resolution in specific atmospheric windows. Our proposed survey, conducted within a single space mission, will be as sensitive as ten CMB-S4 experiments. The survey will complement the frequency coverage in the gaps between atmospheric windows, and make
observations above 300\,GHz with angular resolution matching that of the ground experiments. This complementarity will be key to separate the mixture of SZ effects and CIB emission into their individual contributions.

$\bullet$ Data from other cosmological probes, most notably LSST, will tighten constraints on the cosmological model; {\it Consistency tests using data from different probes are essential for establishing confidence in a cosmological model.} Complementary probes also help lift parameter degeneracies. 

Our proposed survey builds on the technological developments currently ongoing for the next generation experiments in CMB observations and line intensity mapping.

\subsection{Technology challenges and readiness}

The proposed survey does not require the development of technologies or techniques that have not been already demonstrated at some level, either in space, or in ground-based experiments. The main challenges are to scale up existing capability (in particular the number of detectors and frequency range of operation), to demonstrate flight readiness for all components, and to provide the scientific payload with all the necessary resources, in particular in terms of cooling of the telescope and of the focal plane, on-board power, and telecommunication for data download from an orbit around L2.

\noindent\myuline{\it A 4-m class telescope at 8\,K:}\quad 
The baseline telescope has a 3.5-m aperture, and is actively cooled to $\simeq 8\,$K. It is the same size used by {\it Herschel}, albeit much colder. One option proposed for the {\it Origins Space Telescope\/} \citep{origins}, a mission proposed in the US for implementation in the 2020s, has a 5.9-m telescope cooled to 4\,K. 
%
%
For optimal science, a 20\% increase of the aperture size with respect to the baseline (4.2\,m instead of 3.5\,m) would be desirable if the capacity of the fairing of a next-generation European launcher allows it. Launchers with an 8-m fairing, as foreseen for the launch of the {\it Origins Space Telescope\/} flagship mission, are available in the US. 
Although not desirable, the aperture size could be reduced by 20\% (2.8\,m), for an angular resolution of $1.4^\prime$ at 300\,GHz. For comparison, the {\it SPICA\/} space mission, currently in phase A, is planned to have a 2.5-m primary actively cooled to $<10$\,K. 

\noindent\myuline{\it Optical components:}\quad Optical components for mm and sub-mm astronomy are widely available today. Devices and modelling tools are already reliable for various optical configurations (reflective or refractive). Polarization filtering and modulation, spectral filtering, phase control, and stray-light suppression are well understood and can be implemented in a variety of ways, also at cryogenic temperatures. For instruments to be launched post-2035, we can expect reduction in mass and increased compactness.
Solutions based on artificial materials (metamaterials, photonic crystals) are progressing rapidly and will soon offer new tools for astronomy including flat lenses (\citep{Aieta1342} and \citep{Pisano13}) tailored emissivity mm-wave radiators. These developments are interesting for the scientific program in this white paper.

\noindent\myuline{\it Focal plane arrays:}\quad The polarimetric imager uses 3-color pixels with TES bolometers at frequencies below 450\,GHz. The technology is extensively used on the ground over a narrower range of frequencies, and with higher optical loading. TES bolometers have been used aboard balloon instruments starting with the EBEX experiment in 2009~\citep{ebex18}. A flight of {\it LiteBIRD\/} will elevate the TRL of multi-color pixels with TES bolometers to space-flight worthiness. 
The technical milestones for the filter-bank spectrometer are to: (1) demonstrate operation over a broader range of frequencies than have been used to date, which is 330--380\,GHz \cite{Endo2019}; and (2) optimize the operation of the detectors to space loading. 

TES and KID readout technologies are progressing rapidly. By the 2030s we should expect multiplexing factors in the several thousands to have been thoroughly tested. Reduction in power consumption by FPGAs, ADCs, and DACs would further simplify the design of a 2030s survey. 


\newpage
{\small
\bibliography{biblio}
}

\end{document}